# The Effect of Tetrahedral versus Octahedral Network-Blocking Atom Substitutions on Lithium Ion Conduction in LLZO Garnet

James R. Rustad

Corning Incorporated, Corning, NY 14830; present address: University of California, Davis, One Shields Avenue, Davis, CA 95616

$Li_7La_3Zr_2O_{12}$ (LLZO) crystallizes in the garnet structure and has been proposed for use as a solid electrolyte membrane in lithium-based batteries[1]. Pure LLZO is highly conductive at sufficiently high temperature, due to the continuous diamond-structure network of lithium ions which characterizes the LLZO garnet. At low temperature, the lithium ion distribution becomes ordered and the cubic high temperature structure converts to a tetragonal structure with much lower conductivity[2]. The ordering transition can be inhibited by reducing the lithium ion concentration in the conduction network[3,4]. This can be done either by substituting a $5^+$ or $6^+$ ion such as, for example, niobium, tantalum, or tungsten at the $Zr^{4+}$ site, or by substituting a 2+ or 3+ cation, such as $Al^{3+}$, into the tetrahedral network of lithium ions[5,6].

It has not been generally appreciated that, in the latter strategy, the highly charged cation will be much less mobile than the lithium ions and must, to some extent, block the conduction network. Understanding the details of how this network blockage occurs gives some insight into how to optimize the occupation of the conduction network to achieve maximum ionic conductivity. The lithium ion conduction network has essentially two sites: a tetrahedrally-coordinated site at the "nodes" (24d in space group I a -3 d) and an octahedrally-coordinated site (48g) at the "bridge" between the two nodes (in detail the 48g sites sit in the middle of two 96h sites; it is these that are occupied, however since both 96h sites associated with a given 48g site are presumably rarely occupied together, it is easier to discuss the structure in terms of the 48g sites). The conduction network is illustrated in Figure 1. It is possible to choose substituents that would have different preferences between the two sites and an interesting question is to what extent the choice would affect the ionic conductivity in the cubic structure. Just by inspection, it seems clear that, all else being equal, it would be better to block conduction along a "bridge" (the 48g site) than a "node" (the 24d site), just as it's better to block traffic on a single road than at an intersection. But it is not clear exactly how big an advantage might be gained at a given concentration of dopant. The question also arises whether tetrahedral vs. octahedral occupation of substituent hypervalent ions in the lithium conduction network would affect the relative stabilities of the cubic and tetragonal phases. If it did turn out, for example, that octahedral occupation of the hypervalent ions resulted in noticeably higher ionic conductivity in the cubic phase, it might also turn out that octahedral occupation strongly favored the tetragonal phase thermodynamically, making it impossible to produce the cubic phase. This paper addresses the question of how the site preference of a hypervalent ion (24d versus 48g) affects the ionic conductivity of the LLZO for cases in which the substituted ion resides in the lithium ion conduction network. It also addresses the issue of how 24d versus 48g occupation would be expected to influence the energy difference between the tetragonal and cubic phases.

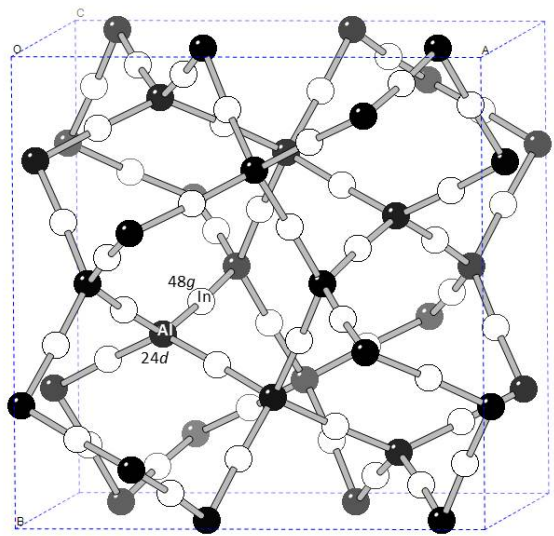

**Figure 1.** Network of 24d (black) and 48g (white) sites in the lithium ion conduction network in garnet-LLZO.

A series of molecular dynamics calculations was performed on aluminum- and indium-substituted cubic LLZO using a modified PMMCS force field[7] (see Supporting Information for specific modifications and simulation protocol). Activation energies and conductivities at 300K were determined using the standard log D (D in cm$^2$/s) versus 1000/T (T in K) plots. The cell size as a function of temperature was estimated using the thermal expansion data in Ref. 8 (which are for pure LLZO; possible effects of Al/In doping on the thermal expansion were not accounted for here). Having obtained the diffusion coefficient at 300K by extrapolating linearly in the activation energy, the electrical conductivity at 300K was estimated from the Nernst Einstein relation:

$$\sigma(\text{ohm}^{-1}\,\text{cm}^{-1}) = z^2 e^2 c \mu_{300}/100 \quad (1)$$

where $z$ is the charge of the migrating ion (+1 in this case), $c$ is the concentration of migrating ions (56 (or 50)/$a^3$(Å$^3$)×10$^{30}$ ions/m$^3$), $e$ is the electric charge 1.602×10$^{-19}$ C, $\mu_{300}$ (s/kg) is the mobility = $D_{300}/kT$, where $D_{300}$ is the lithium ion diffusivity at 300K, and the 100 is to convert the m$^{-1}$ to cm$^{-1}$. To keep the comparisons as simple as possible, the mass of the "indium" ion is chosen to be the same as that of the aluminum ion. This is in keeping with the generic question addressed here about whether it is better to have an immobile ion at the 24d site in the lithium ion conducting network or whether it is better to have an immobile ion at the 48g site. Possible dynamical effects associated with having different ion *masses* at these sites were not explored.

The entire protocol was run three times, with each trial starting from a new set of initial positions obtained by heating the system to 3500 K for 1 nanosecond to erase memory of the previous run. The high-temperature simulation was carried out with Ga$^{3+}$ parameters which resulted in considerable diffusion of the trivalent doping ions and significant occupancy of both 24d and 48g sites. The final configuration of this run was then used as input for the 2000K production run, during which the indium /aluminum atoms quickly settled in to the 48g and 24d sites, respectively. The strong affinity of Al$^{3+}$ for the 24d sites is in agreement with recent first-principles calculations[9].

The results of the molecular dynamics calculations are given in Table 2 and the data are plotted in Figure 2. The raw data are given in the Supporting Information.

**Table 2.** Conductivities (in units of 10$^{-6}$/ohm/cm and activation energies (eV) calculated from molecular dynamics simulations.

|  | Trial 1 | | Trial 2 | | Trial 3 | |
| --- | --- | --- | --- | --- | --- | --- |
|  | σ | E$_a$ | σ | E$_a$ | σ | E$_a$ |
| Al-doped | 8.8 | 0.397 | 6.0 | 0.409 | 2.5 | 0.434 |
| In-doped | 89 | 0.330 | 96 | 0.327 | 52 | 0.348 |

All three trials give the same essential result. The average of the three trials gives 5.8 ×10$^{-6}$ ohm$^{-1}$ cm$^{-1}$ for the conductivity of Li$_{6.25}$Al$_{0.25}$La$_3$Zr$_2$O$_{12}$ and 79 ×10$^{-6}$ ohm$^{-1}$ cm$^{-1}$ for the conductivity of Li$_{6.25}$In$_{0.25}$La$_3$Zr$_2$O$_{12}$, along with an unambiguous drop in the activation energy from 0.413 eV for the aluminum-substituted compound and 0.335 eV for the indium-substituted compound.

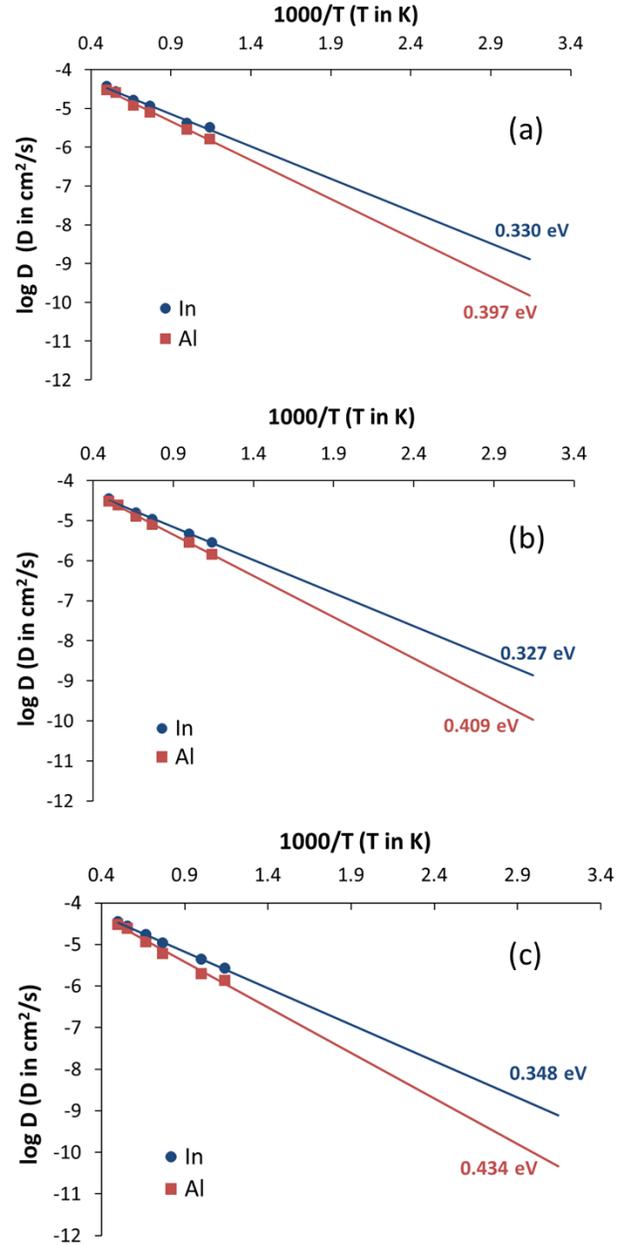

**Figure 2.** log D versus 1000/T plots for trials 1 (a), 2 (b), and 3(c). In-doped composition shown in blue, Al-doped composition shown in red.

Calculations of the tetragonal versus cubic stability are done with the Vienna Ab initio Simulation Program (VASP)[10-13]. The calculations are done with the projector augmented wave method[14,15] at the gamma point (with O, Al, Li$_{sv}$, In$_d$, La, Zr potentials) using the PBE exchange-correlation functional[16,17]. The energy cutoff was 520 eV. Calculations on both cubic and tetragonal phases were done with the measured lattice constants (13.000 Å for cubic and 13.134 Å (a), and 12.663 Å (c) for the tetragonal phase). For each configuration (except for the ordered pure tetragonal LLZO phase, 100 random configurations were generated and optimized to find low-energy configurations for analysis. A sample VASP INCAR file, as well as the optimized lowest-energy configurations (in POSCAR format) for tetragonal and cubic phases for the pure, aluminum-substituted, and indium-substituted compounds are given in the Supporting Information.

Results of the cubic/tetragonal energy differences for the $Li_7La_3Zr_2O_{12}$, $Li_{6.25}Al_{0.25}La_3Zr_2O_{12}$, and $Li_{6.25}In_{0.25}La_3Zr_2O_{12}$ phases are given in Table 3. The calculations indicate that there is no essential difference in tetragonal versus cubic stability between $Li_{6.25}Al_{0.25}La_3Zr_2O_{12}$ and $Li_{6.25}In_{0.25}La_3Zr_2O_{12}$. For both substituted phases, the cubic phase is lower in electronic energy than the tetragonal phase – opposite to what is seen in pure $Li_7La_3Zr_2O_{12}$. This is in qualitative agreement with the calculations on lithium-deficient LLZO in Ref. 4, in which the calculations of cubic versus tetragonal stability were done for a uniform positive charge-compensating background rather than by explicit inclusion of a hypervalent ion into the lithium conduction network. The calculations shown in Table 3 suggest that $Li_{6.25}In_{0.25}La_3Zr_2O_{12}$ can, in fact, be synthesized.

**Table 3.** Electronic energy difference $\Delta E = E_{tet} - E_{cub}$ between cubic and tetragonal phases (kJ/mol)

|  | LLZO | $Li_{6.25}Al_{0.25}La_3Zr_2O_{12}$ | $Li_{6.25}In_{0.25}La_3Zr_2O_{12}$ |
|---|---|---|---|
| $\Delta E^*$ | -8.9 | +4.2 | +3.4 |

*negative number indicates tetragonal phase is more stable

The molecular dynamics calculations presented here show that the tendency of $Al^{3+}$ to occupy the 24$d$ sites in LLZO lithium ion conductors is expected to negatively impact ionic conductivity. The room-temperature ionic conductivity of $Li_{6.25}In_{0.25}La_3Zr_2O_{12}$, in which the $In^{3+}$ resides at the 48g sites, is predicted to be an order of magnitude higher than $Li_{6.25}Al_{0.25}La_3Zr_2O_{12}$. Consideration of the simple tetrahedral topology of the lithium ion conduction network suggests that the increased conductivity arises because the 48g site sits at a "bridge" in the lithium ion conduction network. An immobile ion residing in this location blocks fewer lithium conduction pathways than an ion, such as $Al^{3+}$, that preferentially occupies the 24d site located at a "node" within the network. Other candidates for substituents that might preferentially reside in the 48g sites would be ions with significant octahedral preference due to crystal field effects such as $Cr^{3+}$ (although $Cr^{3+}$, for example, would be less stable toward reduction than $In^{3+}$). $Sc^{3+}$, $Y^{3+}$ or the heavy lanthanide elements could also be used, however, $Y^{3+}$ and the heavy lanthanides are probably large enough to have the possibility of occupying the $La^{3+}$ site without having any beneficial effect on the lithium stoichiometry.

In addition, the electronic structure calculations presented here indicate no large difference in the stabilities of tetragonal and cubic phases for the aluminum-substituted and indium-substituted compounds. In both cases, the cubic phase is more stable than the tetragonal phase. This indicates that it should be possible to make a cubic $Li_{6.25}In_{0.25}La_3Zr_2O_{12}$ phase with significant conductivity enhancement over the cubic $Li_{6.25}Al_{0.25}La_3Zr_2O_{12}$ phase.

## ASSOCIATED CONTENT

**Supporting Information**. Modifications of the PMMCS potential functions; Details of simulation protocol; Tables containing raw data for molecular dynamics calculation of conductivity; VASP INCAR and POSCAR files for $Li_{6.25}Al_{0.25}La_3Zr_2O_{12}$ and $Li_{6.25}In_{0.25}La_3Zr_2O_{12}$

## AUTHOR INFORMATION

### Corresponding Author

* (J.R.R.) James R. Rustad, rustadjr@corning.com

Supplemental Information.

I. Modifications to the PMMCS force field

The modifications/additions to PMMCS were as follows: (1) $La^{1.8+}$-$O^{-1.2}$ parameters were obtained from the $Nd^{1.8+}$-$O^{-1.2}$ parameters by increasing the $r_0$ parameter slightly to give good agreement with the lattice constant of cubic $La_2O_3$; (2) parameters were developed for $In^{1.8+}$-$O^{-1.2}$ (and also $Ga^{1.8+}$-$O^{-1.2}$, see below) by fitting to the structure and elastic properties of $In_2O_3$ and $Ga_2O_3$ (Work on In-O and Ga-O interactions was done by TJ Mustard during internship at Corning Incorporated during summer 2012); (3) the $r_0$ parameter for the $Li^{0.6+}$-$O^{-1.2}$ interaction was increased from 2.681360 Å to 2.75 Å, as this gave better agreement with the experimentally-determined ionic conductivity. The added parameters are given in Table 1.

**Table 1.** Parameters added/modified from the PMMCS force field.

| | $D_{ij}$ (eV) | $a_{ij}$ (Å$^{-1}$) | $r_0$ (Å) | $C_{ij}$ (eV Å$^{12}$) |
|---|---|---|---|---|
| $La^{1.8+}$-$O^{-1.2}$ | 0.014580 | 1.8251 | 3.46002 | 3.0 |
| $In^{1.8+}$-$O^{-1.2}$ | 0.072974 | 1.9342 | 2.82499 | 5.0 |
| $Ga^{1.8+}$-$O^{-1.2}$ | 0.038263 | 2.1052 | 2.6681 | 0.9 |
| $Li^{0.6+}$-$O^{-1.2}$ | 0.001114 | 3.429506 | 2.75 | 1.0 |

II. Simulation Protocol

The calculations were done using a code developed by JW Halley, JR Rustad, E Wasserman, and A Rahman. Program units were length= Ångstroms, mass=AMU, energy = $e^2$/Å. With these choices, the unit of time is fixed at 2.6828×10$^{-15}$ sec. The cell size as a function of temperature was estimated using the thermal expansion data in Ref. 8 (which are for pure LLZO; possible effects of Al/In doping on the thermal expansion were not accounted for here). The calculations were run with a single Nose-Hoover thermostat with a mass parameter of 500 in program units and a time step of 0.05 in program units. The fifth-order Gear algorithm was used to integrate the equations of motion for the atoms. The Ewald method was used to compute energies and forces with a real space cutoff of 1.5*a* (one real space shell around the central unit cell) and an Ewald sum convergence parameter $\alpha L$=6.0.

Statistics on the mean-square displacement of the lithium, aluminum, and indium ions were collected over 4×10$^6$ time steps (537 picoseconds), keeping track of the mean square displacements up to 15 ps. The slope of the mean-square displacement versus time graph was computed from 2-15 picoseconds. Runs were carried out at 2000K (*a*=13.335 Å), 1800K (*a*=13.292 Å), 1500K (*a* =13.227 Å), 1300K (*a*=13.184 Å), 1000K (*a*=13.120 Å), and 875K (*a*=13.093 Å).

III. Data from molecular dynamics calculations

Table S1. Results for Al-LLZO (trial 1)

| T (K) | a (Å) | MSD slope | D cm$^2$/s | log D | linear fit† | D cm$^2$/s (fitted) | σ (ohm$^{-1}$ cm$^{-1}$) ‡ |
|---|---|---|---|---|---|---|---|
| 300 | 12.970 | | | | -10.20933 | 6.18E-11 | 8.77E-06 |
| 875 | 13.093 | 0.096 | 1.60E-06 | -5.79722 | -5.82794 | 1.49E-06 | 2.05E-01 |
| 1000 | 13.120 | 0.168 | 2.80E-06 | -5.55288 | -5.54220 | 2.87E-06 | 3.94E-01 |
| 1300 | 13.184 | 0.467 | 7.79E-06 | -5.10858 | -5.08062 | 8.31E-06 | 1.12E+00 |
| 1500 | 13.227 | 0.699 | 1.17E-05 | -4.93367 | -4.87547 | 1.33E-05 | 1.78E+00 |
| 1800 | 13.292 | 1.505 | 2.51E-05 | -4.60051 | -4.65322 | 2.22E-05 | 2.93E+00 |
| 2000 | 13.335 | 1.778 | 2.96E-05 | -4.52832 | -4.54210 | 2.87E-05 | 3.75E+00 |

†-2.0002*(1000/T)-3.5420
‡D/10000*50/($a^3$*1E-30)*(1.602E-19$^2$)/(1.3806E-23*300)/100

Table S2. Results for Al-LLZO (trial 2)

| T (K) | a (Å) | MSD slope | D cm$^2$/s | log D | linear fit† | D cm$^2$/s (fitted) | σ (ohm$^{-1}$ cm$^{-1}$) ‡ |
|---|---|---|---|---|---|---|---|
| 300 | 12.970 | | | | -10.37133 | 4.25E-11 | 6.04E-06 |
| 875 | 13.093 | 0.085 | 1.42E-06 | -5.84749 | -5.85523 | 1.40E-06 | 1.93E-01 |
| 1000 | 13.120 | 0.168 | 2.81E-06 | -5.55164 | -5.56070 | 2.75E-06 | 3.77E-01 |
| 1300 | 13.184 | 0.469 | 7.81E-06 | -5.10735 | -5.08492 | 8.22E-06 | 1.11E+00 |
| 1500 | 13.227 | 0.749 | 1.25E-05 | -4.90386 | -4.87347 | 1.34E-05 | 1.79E+00 |
| 1800 | 13.292 | 1.450 | 2.42E-05 | -4.61686 | -4.64439 | 2.27E-05 | 2.99E+00 |
| 2000 | 13.335 | 1.805 | 3.01E-05 | -4.52161 | -4.52985 | 2.95E-05 | 3.86E+00 |

†-2.0617*(1000/T)-3.4990
‡D/10000*50/($a^3$*1E-30)*(1.602E-19$^2$)/(1.3806E-23*300)/100

Table S3. Results for Al-LLZO (trial 3)

| T (K) | a (Å) | MSD slope | D cm$^2$/s | log D | linear fit† | D cm$^2$/s (fitted) | s (ohm$^{-1}$ cm$^{-1}$) ‡ |
|---|---|---|---|---|---|---|---|
| 300 | 12.970 | | | | -10.74780 | 1.79E-11 | 2.54E-06 |
| 875 | 13.093 | 0.081 | 1.35E-06 | -5.86884 | -5.95460 | 1.11E-06 | 1.53E-01 |
| 1000 | 13.120 | 0.118 | 1.96E-06 | -5.70684 | -5.64200 | 2.28E-06 | 3.13E-01 |
| 1300 | 13.184 | 0.363 | 6.05E-06 | -5.21847 | -5.13703 | 7.29E-06 | 9.86E-01 |
| 1500 | 13.227 | 0.696 | 1.16E-05 | -4.93527 | -4.91260 | 1.22E-05 | 1.64E+00 |
| 1800 | 13.292 | 1.463 | 2.44E-05 | -4.61303 | -4.66947 | 2.14E-05 | 2.82E+00 |
| 2000 | 13.335 | 1.809 | 3.02E-05 | -4.52068 | -4.54790 | 2.83E-05 | 3.70E+00 |

†-2.1882*(1000/T)-3.4538
‡D/10000*50/($a^3$*1E-30)*(1.602E-19$^2$)/(1.3806E-23*300)/100

Table S6. Results for In-LLZO (trial 1)

| T (K) | a (Å) | MSD slope | D cm$^2$/s | log D | linear fit† | D cm$^2$/s (fitted) | σ (ohm$^{-1}$ cm$^{-1}$) ‡ |
|---|---|---|---|---|---|---|---|
| 300 | 12.970 | | | | -9.20187 | 6.28E-10 | 8.92E-05 |
| 875 | 13.093 | 0.195 | 3.26E-06 | -5.48742 | -5.55494 | 2.79E-06 | 3.85E-01 |
| 1000 | 13.120 | 0.247 | 4.11E-06 | -5.38615 | -5.31710 | 4.82E-06 | 6.61E-01 |
| 1300 | 13.184 | 0.680 | 1.13E-05 | -4.94557 | -4.93289 | 1.17E-05 | 1.58E+00 |
| 1500 | 13.227 | 0.959 | 1.60E-05 | -4.79638 | -4.76213 | 1.73E-05 | 2.32E+00 |
| 1800 | 13.292 | 1.601 | 2.67E-05 | -4.57369 | -4.57714 | 2.65E-05 | 3.49E+00 |
| 2000 | 13.335 | 2.179 | 3.63E-05 | -4.43982 | -4.48465 | 3.28E-05 | 4.28E+00 |

†-1.6649*(1000/T)-3.6522
‡D/10000*50/($a^3$*1E-30)*(1.602E-19$^2$)/(1.3806E-23*300)/100

Table S7. Results for In-LLZO (trial 2)

| T (K) | a (Å) | MSD slope | D cm$^2$/s | log D | linear fit† | D cm$^2$/s (fitted) | σ (ohm$^{-1}$ cm$^{-1}$) ‡ |
|---|---|---|---|---|---|---|---|
| 300 | 12.970 | | | | -9.17177 | 6.73E-10 | 9.56E-05 |
| 875 | 13.093 | 0.171 | 2.85E-06 | -5.54442 | -5.56296 | 2.74E-06 | 3.78E-01 |
| 1000 | 13.120 | 0.280 | 4.67E-06 | -5.33073 | -5.32760 | 4.70E-06 | 6.45E-01 |
| 1300 | 13.184 | 0.648 | 1.08E-05 | -4.96646 | -4.94741 | 1.13E-05 | 1.53E+00 |
| 1500 | 13.227 | 0.948 | 1.58E-05 | -4.80136 | -4.77843 | 1.67E-05 | 2.23E+00 |
| 1800 | 13.292 | 1.429 | 2.38E-05 | -4.62317 | -4.59538 | 2.54E-05 | 3.35E+00 |
| 2000 | 13.335 | 2.129 | 3.55E-05 | -4.44995 | -4.50385 | 3.13E-05 | 4.10E+00 |

† -1.6475*(1000/T)-3.6801
‡ D/10000*50/($a^3$*1E-30)*(1.602E-19$^2$)/(1.3806E-23*300)/100

Table S8. Results for In-LLZO (trial 3)

| T (K) | a (Å) | MSD slope | D cm$^2$/s | log D | linear fit† | D cm$^2$/s (fitted) | σ (ohm$^{-1}$ cm$^{-1}$) ‡ |
|---|---|---|---|---|---|---|---|
| 300 | 12.970 | | | | -9.43587 | 3.67E-10 | 5.20E-05 |
| 875 | 13.093 | 0.158 | 2.64E-06 | -5.57919 | -5.59903 | 2.52E-06 | 3.48E-01 |
| 1000 | 13.120 | 0.262 | 4.37E-06 | -5.35946 | -5.34880 | 4.48E-06 | 6.14E-01 |
| 1300 | 13.184 | 0.637 | 1.06E-05 | -4.97383 | -4.94458 | 1.14E-05 | 1.54E+00 |
| 1500 | 13.227 | 1.037 | 1.73E-05 | -4.76238 | -4.76493 | 1.72E-05 | 2.30E+00 |
| 1800 | 13.292 | 1.616 | 2.69E-05 | -4.56964 | -4.57031 | 2.69E-05 | 3.55E+00 |
| 2000 | 13.335 | 2.099 | 3.50E-05 | -4.45615 | -4.47300 | 3.37E-05 | 4.40E+00 |

† -1.7516*(1000/T)-3.5972
‡ D/10000*50/($a^3$*1E-30)*(1.602E-19$^2$)/(1.3806E-23*300)/100

## IV. Input and configuration files for VASP calculations

INCAR file

```
ENCUT=520
ISMEAR=0
SIGMA = 0.2
IBRION=2
ISIF=0
LREAL=Auto
LCHARG=.FALSE.
LWAVE=.FALSE.
NSW = 200
NPAR=4
```

CONTCAR files:

CONTCAR_Al_cub_006
```
llzo
 1.00000000000000000
    13.0000000000000000     0.0000000000000000     0.0000000000000000
     0.0000000000000000    13.0000000000000000     0.0000000000000000
     0.0000000000000000     0.0000000000000000    13.0000000000000000
   Zr   O   La   Al   Li
   16   96   24    2   50
Direct
   0.7495354620600385  0.2472668163187947  0.7485632078142989
   0.2531961613553400  0.7502716350839108  0.2502738776493690
   0.7498086742436334  0.7491280749974174  0.2495216965281800
   0.2479509202289512  0.2508684193666415  0.7482732044888374
   0.2487964532283446  0.7506948281104182  0.7456722150747147
   0.7504258688157606  0.2512195718997059  0.2480476886327961
   0.2523840677389417  0.2488176605300353  0.2476902363927323
   0.7507458762048843  0.7487964013706943  0.7471838984734634
   0.4965316024142467  0.5031518565722994 -0.0017496287080125
  -0.0024815745480275  0.0052395665943314  0.4959900500726152
   0.5059960372955498  0.0081070247297182  0.5029531562075515
   0.0035457216788258  0.5055241328347569 -0.0008393052835226
   0.0046211484997031  0.5073909333489046  0.4972299469948950
   0.5003855471380195  0.0048999109247658 -0.0037749202203160
  -0.0010210713734865  0.0039381558024040  0.0014729040019511
   0.4985549185467750  0.5031296966133908  0.5025209385274487
   0.5510888932979973  0.3523854711255614  0.0326458561351023
   0.0519218137491239  0.8557430964958640  0.5325186581521718
   0.9383543395357323  0.6516345551373015  0.5358415252440883
   0.4423546022842675  0.1518668176676089  0.0359944917298291
   0.4530001745348952  0.8573582959701559  0.4631649529704735
   0.9512848031565488  0.3532025363997325  0.9661683176972943
   0.0562297491758759  0.1513219551271018  0.9618167622845055
   0.5618671421931585  0.6498870807971919  0.4618123468063183
   0.3537221897932213  0.0252444734283117  0.5578972840675116
   0.8528165024957820  0.5382060728912221  0.0544037122357623
   0.6477618310334511  0.5378073565711449  0.9424665503169198
   0.1482577359065530  0.0272325135277717  0.4373298906656417
   0.8591382710730555  0.4652515451013720  0.4385392384682080
   0.3538486878041046  0.9648053555768416  0.9441284504346436
   0.1457810197977970  0.9640941285739903  0.0538648645864403
   0.6501071208053552  0.4673179863992513  0.5528148529114479
   0.0269310509320167  0.5678772267975366  0.3479053136213751
   0.5242264116684966  0.0610446571518011  0.8492393835846225
   0.5423516633775337  0.9429955064090830  0.6485568438207066
   0.0390274517663518  0.4505025657737493  0.1491916183864054
   0.4642240086125007  0.4484219446078352  0.8506986644614506
   0.9600167994910289  0.9468244786667172  0.3467187378670570
   0.9762886847257257  0.0639636400612336  0.1500983093146709
   0.4752016468981887  0.5651569222577264  0.6502688894582391
   0.9438444874241679  0.1600985118084283  0.4734548525082878
   0.4459746434156808  0.6586364670794184  0.9686613043317414
   0.5569403778382460  0.8519525174941140  0.9668163903048659
   0.0596743819030862  0.3565898465319894  0.4712310071453651
   0.0538346478917536  0.6599921597652871  0.0274824078633587
   0.5616706185928403  0.1605055227623044  0.5238109326308039
   0.4474667397739556  0.3521820615238039  0.5254703211203123
   0.9436191481415300  0.8520266647961855  0.0311160969381365
   0.1543390741807003  0.4688617156211649  0.9474694720224166
   0.6569159834606387  0.9728898723351687  0.4444893071555115
```

```
 0.8463151376856299   0.9706845781400478   0.5495970738275676
 0.3471370628181818   0.4684596659276510   0.0563235535362525
 0.6552799746245871   0.0434960401444798   0.0498924841939796
 0.1582310463392656   0.5420602557880457   0.5482321743878055
 0.3442424529072564   0.5392970470076517   0.4496971774456334
 0.8442632946789462   0.0419273623971692   0.9489255444103266
 0.4777213774099339   0.9521014826700475   0.1531504160956629
 0.9741779978117709   0.4536910033361787   0.6497680661319516
 0.9720562104087452   0.5524305503688824   0.8465895011120167
 0.4704874293612367   0.0566184003889941   0.3489549321568069
 0.0296630163952704   0.0532512812434089   0.6490374446106048
 0.5292976587407814   0.5527556215719832   0.1515207712674401
 0.5256355105998584   0.4538259466263542   0.3464971215143606
 0.0243417582144245   0.9513363291372892   0.8468335221608655
 0.3060911839362183   0.7910108441007958   0.1018940182062577
 0.7977948183794116   0.2951300108901189   0.6021399300438575
 0.6958637256427281   0.2956710455033922   0.3930961805992054
 0.1948257256377114   0.7903312695309160   0.8943089783708260
 0.1906210336269587   0.2125769891672334   0.5977686770805180
 0.6916594216265898   0.7164728104974514   0.0989526396782864
 0.8093972089161825   0.7167433411855701   0.8976661486760611
 0.3010604447433893   0.2141203851505298   0.4012799453643194
 0.0973306510631727   0.3090012477816737   0.7795445735570145
 0.6002491451549091   0.8039538833533821   0.2856113888522260
 0.4010815100208610   0.6905177699644606   0.2733936791435544
 0.9003977684486528   0.1917657488356005   0.7717098624692057
 0.6034177373245172   0.1904915022941872   0.2238811803160859
 0.0998081536669178   0.6914719060204981   0.7232208756678570
 0.9018542024564205   0.8048902689338868   0.7106890130402216
 0.4006425244224876   0.3077580859044314   0.2212238405406152
 0.7881433888888757   0.1071829881661226   0.3096346344654749
 0.2866216698950116   0.6048631825594808   0.8078032950206979
 0.2726868153812069   0.3986819102023204   0.6878622826124818
 0.7833576904277335   0.8985040886450191   0.1933076278766575
 0.2134621700100014   0.6060037433700829   0.1871416781600454
 0.7252282685118423   0.1019500898292296   0.6854584162701907
 0.7174177058860597   0.8985581936720732   0.8035364761757698
 0.2157235989976724   0.3976573377034491   0.3022874553947596
 0.7000456925736978   0.2160191742539702   0.9056111709679444
 0.2015932197084989   0.7167039152436117   0.4046084893802226
 0.3011597670554183   0.7176505214835411   0.5905891796229639
 0.7990301123996739   0.2180092687368231   0.0901913732492323
 0.8020440509887098   0.7788572745875453   0.4012926561133586
 0.3069197168891988   0.2734277685866686   0.8983119712464274
 0.1906851688364293   0.2738794384289884   0.0983172924210583
 0.7007153283368329   0.7783717721176542   0.5945202847879604
 0.9031774908242571   0.7013051297684115   0.2170126014289881
 0.3929999156235510   0.1899227572462047   0.7151169256866273
 0.5990439806385293   0.3064159212895920   0.7204191183174899
 0.0980064552959154   0.8040956958154027   0.2237283223334938
 0.4027249136153102   0.8030764432527434   0.7726772353363961
 0.9060061376973936   0.3044520405926216   0.2747986501459827
 0.1002625768812717   0.1971341443299759   0.2870148164999874
 0.5982113342197529   0.7016671389631727   0.7788197733189147
 0.2170001840442380   0.9025923750691779   0.6919587427746313
 0.7183228173050950   0.4032492752731368   0.1900404003595628
 0.7211576557379390   0.5941418296104156   0.2994057293695692
 0.2159208593873190   0.0951945585511785   0.8007828193520359
 0.7842338318120838   0.4004343948253982   0.8078990495922082
 0.2843407250600817   0.9021183652910877   0.3040929129983959
 0.2820879411861736   0.0950574549586663   0.1975365074826861
 0.7799082027625035   0.5933785591814301   0.6983406447431304
 0.4943468140798244   0.2506564948872760   0.8630561544323926
-0.0004300547482543   0.7522370520153306   0.3811283460245044
 0.5016937963179747   0.7517075597518420   0.6152582685581079
 0.0049099691965805   0.2506988358105469   0.1234540590061583
 0.2515175876862457   0.8821298692674523   0.4980464924217594
 0.7496601617832725   0.3829155843262922  -0.0027196904045027
 0.7501751916789382   0.6203426214181270   0.4984503476015387
 0.2500819290063777   0.1250812160958449  -0.0029438650499575
 0.8837051728531983   0.5021970876409768   0.2478910464306062
 0.3813937424200173   0.0030110641599348   0.7452408356405944
 0.6166527176157737   0.4997828895036461   0.7493728814688198
 0.1213316477629706   0.0000571353903720   0.2503098634670249
 0.0028686268988847   0.2540172235191950   0.6268337886560672
 0.5011935150017699   0.7536549490390373   0.1303193615292241
 0.0004507779815849   0.7547716259928574   0.8661982709909422
 0.5000059147448276   0.2539887705272571   0.3728211645981578
```

```
  0.2514439815447290    0.6342955915381715   -0.0022890378261599
  0.7510122957425246    0.1344778566218695    0.4991740208095161
  0.7492229891131372    0.8735827030825170   -0.0026726302623835
  0.2532903371064676    0.3766562348947827    0.5025715645843748
  0.6331933913242936    0.0046175291831441    0.2516914519115733
  0.1344293877671447    0.5043380167994211    0.7487025068684919
  0.8683799248753358    0.0027054161342358    0.7450347013978377
  0.3681849823605461    0.5027558567935116    0.2490112952720055
  0.7500958112103942    0.1275418721124670   -0.0015383056040774
  0.2516403752797545    0.6290610763479398    0.4981281219467675
  0.2952936002164117    0.4133974739869036   -0.0722761952306062
  0.0701167195241487   -0.0900687479298565    0.6787086977342754
  0.6717952764928212    0.4408087843830172    0.3345987624351391
  0.4312219959166026   -0.0871562575521510    0.3180361253490681
 -0.0672153465962047    0.8482850115710409    0.1798608525393581
  0.4876115938884438    0.2439451882238225    0.6257101563440738
  0.4341255927277051    0.6526849573319897    0.8170524008810187
  0.0750674085908149    0.3394205544936688    0.3281870528185103
  0.7575429929123922    0.3821972379040072    0.4923542136733773
  0.2500141772584756    0.8784791356676772   -0.0019600057377654
  0.7497750496637496    0.6275827658350422   -0.0021370449390893
  0.8122623655987370   -0.0724159889197070    0.4041001483559689
  0.5693193445841671   -0.1478147189689460    0.8155396923526498
  0.4856903802838838    0.7441338445621953    0.3779878601073799
  0.0686598706875938    0.1057494746304941    0.8116319003010818
  0.1722225593988086    0.9436605457426844    0.8349881221494933
  0.3210602746705956    0.0666188823376118    0.3544295363940496
  0.8258917480836448    0.5583560065305224    0.8413252621187362
  0.8268380086988809    0.4391703442147806    0.6632187594026372
  0.3294932203333717    0.9422009206752531    0.1603728410357540
  0.6750767180479309    0.5602682434090019    0.1552335626407738
  0.1763097706604323    0.0657911295410567    0.6482829655500713
  0.8715593515225692    0.0123519859654711    0.2371576570177966
  0.0638873286872478    0.6586311336938681    0.1770839130999500
  0.9414734442967045    0.2196971084306798    0.9025207923103832
  0.3993985569352022    0.3124361645976751    0.0750974363864478
  0.8511075783329615    0.8220001535800667    0.5633969484589861
  0.6487041258932675    0.8235686543197659    0.4326048059229920
  0.1079200566488588    0.3175546235962620    0.9249245922569673
  0.0156371945336347    0.7442940890913645    0.6189043198610328
  0.5646704706452426    0.2147195977407594    0.0888813459080248
  0.3748078036335645    0.4917633648160648    0.7358658403791019
  0.9496526165620606    0.1630911407545621    0.3279747935077940
  0.1888112229276222    0.4222744533129951    0.0957556763056697
  0.6908459668221653    0.9270606699926743    0.5924124955861735
  0.6255131472561080    0.0205217250837234    0.7520233205833945
  0.1296385288667530    0.5095620301527388    0.2623530137332137
  0.5720525120721023    0.5971337366178434    0.3124392039735975
  0.4308117845222148    0.1052273034119614    0.1875244663056121
  0.9282746454014804    0.5971624942980019    0.6865072567207263
  0.5661464892952386    0.3986225675633630    0.1803960014733703
  0.9374130524158404    0.3970042374447464    0.8160802396787931
  0.6217417710970909    0.3227762341890131    0.5735406859031629
  0.0914311156419240    0.8161346467978007    0.0754372066104348
  0.1489439078123339    0.1733942486294667    0.4368782681677092
  0.5903330389337298    0.6919624970636594    0.9244785471186271
  0.9076037809847290    0.6930631125846206    0.0699918846473481
  0.3176111287947097    0.1603676744959944    0.5419600381105818
  0.4108331449322261    0.8157308120657025    0.9202834225565327
  0.9159266560844914    0.3183909863855113    0.4216337183861160
CONTCAR_Al_tet_090
llzo
 1.00000000000000000
    13.1340000000007997    0.0000000000000000    0.0000000000000000
     0.0000000000000000   13.1340000000007997    0.0000000000000000
     0.0000000000000000    0.0000000000000000   12.6630000000000003
   Zr   O   La   Al   Li
   16   96   24    2   50
Direct
  0.2431603312710625    0.2510152491816933    0.2458926853595741
  0.7387612663007311    0.7515856100856707    0.7498984828347413
  0.7446687799670968    0.2623194212270529    0.7503338001035336
  0.2454203552246596    0.7630458293038168    0.2488344871416731
  0.9985677053204143    0.5030776092579300    0.9999026063502087
  0.4967851008188597    0.0064400495354008    0.5034354655849836
  0.9977158866351016    0.5054775909528021    0.4983775946231667
  0.4954250302424853    0.0079766602976693    0.0038908389354598
```

```
0.2418303537642805   0.2585957901181812   0.7464042794697225
0.7429154089385042   0.7556274372476711   0.2517202098252590
0.7459249606956556   0.2577875541319363   0.2499366500329676
0.2453104034867878   0.7553990106481218   0.7484916299908915
0.9938603832496949   0.0103129052028490   0.5017446504536766
0.4962353031692170   0.5120345256546471   0.9987080019500282
0.9945872273475517   0.0081335259229559   0.0008914111136639
0.4939467178576523   0.5095068178527068   0.5004326102136153
0.2956728039889426   0.2894947470609156   0.4024388672977552
0.7999054677173664   0.7898221353649690   0.9037175813638962
0.6873571365129003   0.2889470568844367   0.9044597029212270
0.1888478451802083   0.7885515149229976   0.4036061262819522
0.9602781129489547   0.4479563635896456   0.1524147379218316
0.4581694758484227   0.9529281343639507   0.6568961254678899
0.0282751166866258   0.4541214371316772   0.6543429302399619
0.5301217018457753   0.9532543191098035   0.1574017537928139
0.6864228060371831   0.7312655512444650   0.5950472630632755
0.1801886034625604   0.2292443113711125   0.0940620922590987
0.3020127486273707   0.7238107648906558   0.0956338597762058
0.8012249203178478   0.2248386623765682   0.5973539738508385
0.0278374428708035   0.5643444315815361   0.8469701706634600
0.5307514173204462   0.0626482175662398   0.3504105332967581
0.9634323155479352   0.5620975304743067   0.3441938985958514
0.4562140764662914   0.0660285856878854   0.8525434451229277
0.2994376147529927   0.2319025661594274   0.8975516544761716
0.8026723355768829   0.7219713688439904   0.4043983286789518
0.6902430183986236   0.2224188225141119   0.4032589388172156
0.1894178884593093   0.7195005720415405   0.9020005195474697
0.9601808586003550   0.0618968410479171   0.6574569447330738
0.4615949857797201   0.5625128720529706   0.1562773625390523
0.0288740995711867   0.0682483667969504   0.1479479222980120
0.5188252062302892   0.5733518134220392   0.6477102891448765
0.6895839081370437   0.7960194995759684   0.0996188583041206
0.1855277953078719   0.2977363001512980   0.5951951718295480
0.3005518878649486   0.7892942576350507   0.5948104661138401
0.7991525960918101   0.2887136642323148   0.0947155639675190
0.0289066041259596   0.9452945421147304   0.3534778905908095
0.5291874111061980   0.4480508984039319   0.8500998748479203
0.9632321535704432   0.9487837687326748   0.8461113711738012
0.4600031684293686   0.4479219768877942   0.3459029482702040
0.0975582010827546   0.2037077659449388   0.7854903956822298
0.5981896071604854   0.7042961999417026   0.2862147065563250
0.8933960680046321   0.2041443296039334   0.2852362897257011
0.3912037648092299   0.7039009039578821   0.7876667160901600
0.0514878033069472   0.6562548012332368   0.5341346286053791
0.5549274343834120   0.1556375468145983   0.0375020320175169
0.9499182826678393   0.6591433332715481   0.0452052221082266
0.4510121513920370   0.1632376344665551   0.5422825155164588
0.8985764338063507   0.8048308491196398   0.2073940302172468
0.3973596621298374   0.3059454185332078   0.7063668912763109
0.0940426840460314   0.8073621762844948   0.7134831964384623
0.5957222842046233   0.3098164327161324   0.2135946635026758
0.9462657538758473   0.3568659037008569   0.4633289350064071
0.4415929901117122   0.8596179260812669   0.9652662428946392
0.0498301434564107   0.3595648252958115   0.9602328907982639
0.5472050499235186   0.8616137325222973   0.4647263069548005
0.0983290402537379   0.3152615433343695   0.2847658310483995
0.5955564524793296   0.8152611345401479   0.7861441376744794
0.8897364565519842   0.3151516988190131   0.7881982508870566
0.3903590568149880   0.8149225655434653   0.2883818286148500
0.0425322523054135   0.8500027056226971   0.0313225565875923
0.5449849162750376   0.3536086501847709   0.5360439935705206
0.9366326495393026   0.8610243390807809   0.5379317406928031
0.4428999456062777   0.3621594900955564   0.0373677346731849
0.8881640463045885   0.6969303697206772   0.7122105836805358
0.3946560947308169   0.1972743656902799   0.2265807864452703
0.0930364017637320   0.7089908165633901   0.2068437498913694
0.5914831662119281   0.2086533676274032   0.7090016128683420
0.9395271853084499   0.1514214980879494   0.9594334870333969
0.4372937242233965   0.6522402332683456   0.4579894809365643
0.0520381066564029   0.1578094397189781   0.4795762883873116
0.5533060833052826   0.6610442244170894   0.9806414354311164
0.2667305163183593   0.1096278410953571   0.6854942898370924
0.7710779475501481   0.6065189477249031   0.1953626860792890
0.7187660881293110   0.1041253593970487   0.2004360189528926
0.2170278736204936   0.6042838039219867   0.6959531011976294
0.1485841602617451   0.4818320944449597   0.4415484120763434
0.6455367638847807   0.9810494361796730   0.9445614505374074
```

```
0.8453132985549163  0.4818760951804262  0.9522813763995777
0.3437046836375697  0.9827477720600493  0.4535674641198634
0.7222946346843313  0.9080441944169162  0.3037013188834688
0.2192042163229219  0.4106810733859760  0.7980183835517645
0.2692788879661118  0.9059707295181854  0.8036283550650244
0.7733571377940618  0.4075916328417215  0.3047879008170105
0.8470755931605500  0.5322459814925495  0.5452300224139774
0.3438399568494149  0.0327504992661926  0.0557628063177288
0.1476094819645265  0.5296130691088266  0.0576899740834621
0.6468568963296127  0.0313542136830909  0.5611011374564410
0.2767066875570079  0.4041257727022303  0.1982834031869145
0.7738376261788745  0.9060835734733254  0.7035458805215358
0.7180996275206911  0.4097365017629431  0.6962802091613082
0.2173704930455891  0.9093238927950800  0.1912977350654526
0.1481853716005777  0.0362976543110083  0.9486922682345809
0.6423004744883527  0.5295434571123931  0.4380975630732017
0.8452485986754841  0.0384507916466403  0.4449501734471537
0.3463013883867533  0.5357130526756274  0.9445823693419347
0.7166265850468040  0.6041006210530426  0.8118852953546208
0.2154539157376644  0.1025363530725573  0.3026439776640590
0.2699014424435464  0.6081861319500248  0.2953589150238351
0.7694634628059933  0.1079339499022171  0.7989498152053998
0.8468023091255126  0.9814162043285780  0.0592769806053896
0.3453219438204794  0.4801964207407829  0.5564239411775004
0.1448427229924555  0.9743684228753805  0.5532681936034224
0.6471358441269228  0.4765206328084599  0.0509890946592474
0.4957105191595785  0.2569026350538316  0.3683775238077919
0.9945534825075698  0.7551999195522637  0.8662110182929604
0.4937447346892170  0.2597624289933758  0.8848313938693221
0.9953152387279416  0.7587887932117032  0.3828903687319754
0.9975564555369414  0.2545976667745123  0.1272488226616548
0.4966460223647816  0.7542031798049520  0.6282213536341303
0.9941104746212652  0.2603430260944761  0.6248368782164261
0.4958161323019951  0.7592624899372842  0.1270386785952253
0.2370646203114637  0.1326302244648007  0.4967152555136310
0.7403996873900003  0.6311185193935612  0.0002723887201047
0.7452630110720786  0.1270480643915763  0.9996806265160459
0.2470291293152631  0.6266361161648464  0.4972992079244832
0.1230160954201514  0.5019790313708103  0.2488925937906932
0.6206825584881374  0.0031370610504526  0.7526226600959768
0.8629506534472008  0.5092617430051363  0.7490046965310275
0.3660741760748539  0.0094393938713779  0.2509839115419412
0.7470783682315498  0.8863829119507872  0.5024282820536180
0.2513259153267609  0.3817758362350363  0.0013648277092628
0.2490426996676213  0.8828351132671523  0.9993674817364154
0.7428275760716114  0.3903527174453541  0.5034095917034129
0.1256297452301403  0.0081496023694610  0.7506714938480334
0.6228222612947751  0.5052475636715078  0.2495754619587454
0.8794174530033486  0.0129185481397548  0.2519436365311064
0.3787688021714329  0.5110861648608346  0.7504334564312004
0.4978466951081215  0.2621749181021296  0.6257919994939503
0.9978375587829085  0.7608917350323187  0.1246503409537012
0.4826938968415593  0.2503788791982484  0.1237151597730709
0.9884157401475974  0.7549907596773144  0.6235599489429634
0.0034393902049513  0.2453559832131322  0.3726293117144555
0.5002009351118418  0.7484533360542286  0.8750219311090385
0.9920010919724257  0.2570954333494443  0.8735365272755841
0.4953519882267952  0.7566595743477217  0.3755491497482369
0.6800893581439130  0.0754186363643427  0.3610576320660649
0.1784473154762958  0.5714716719565381  0.8573953119982518
0.3167192349549517  0.0918845504744985  0.9168464440709245
0.8181073350810650  0.5739747534781421  0.3773094198667187
0.1946582382121180  0.0849223964438220  0.0919920506682382
0.8143480665237059  0.0788117357671905  0.6383503312624960
0.3160587772447430  0.5776892388109375  0.1359269710287463
0.3122801182946299  0.9370821929094875  0.6127093538517180
0.8134563440279757  0.4353273047269789  0.1137907826151725
0.6789460138999860  0.9427542084631253  0.1449192724581289
0.1779673253405231  0.4446917731253676  0.6416078193090580
0.8159229598824240  0.9322734053667862  0.8588818416325701
0.3147691782588083  0.4312768669080205  0.3584040800186806
0.1743187707878588  0.9326688748517609  0.4034202442200296
0.6729866836929110  0.4349914167403323  0.8950265541474515
0.8207536484192181  0.6703741216331630  0.5590082172697113
0.6715670358117981  0.1697363011110957  0.5454729349234647
0.1714993527457820  0.6676187981826889  0.0445269037977655
0.0736963661343545  0.4193424248628724  0.8067811737299286
0.5755737699440003  0.9197762013302878  0.3111748059055935
```

```
  0.9176019307725084    0.4093007839500485    0.3094471260137248
  0.4125507441623832    0.9161885623853906    0.8119286930338085
  0.6536943900871501    0.8387899960011033    0.9450161061891184
  0.1519426771944489    0.3394520580118851    0.4399874784889398
  0.3364328507307258    0.8363516077589968    0.4434336518828487
  0.8319892177213523    0.3371408953188893    0.9443281133020293
  0.9103060705060728    0.5810240508992760    0.2001485745096751
  0.4022145853611370    0.0807752097763949    0.7111671892377142
  0.0728892752076731    0.5984898016349982    0.6888541971982770
  0.5748461789010189    0.0986357324151244    0.1933299541698850
  0.3151358947187036    0.3451227831420536    0.5427417672788185
  0.8185613353043143    0.8447140452315247    0.0468890312976908
  0.6493942576361421    0.3311478669641834    0.0623307642567792
  0.1524337097394770    0.8299735298905827    0.5630041447221282
  0.6189200377320300    0.5164364861934506    0.7359725645134817
  0.9140480918325670    0.0926766657905778    0.8073669145630425
  0.4136740401641840    0.5921675968995640    0.3063919995341408
  0.6611733441970675    0.6709918197596617    0.4522399167138263
  0.1511844350532576    0.1843463944768431    0.9379468895996125
  0.3341885243601077    0.6773475233045546    0.9409624522870415
  0.8353631498362689    0.1821691466931452    0.4431514496752035
  0.9164968908037074    0.9190941809075565    0.6916355290340218
  0.4190582433959499    0.4185794402128908    0.1901250730619898
  0.1152106619367048    0.0056252392381211    0.2466585195528701

CONTCAR_In_cub_075
llzo
   1.000000000000000
    13.0000000000000000    0.0000000000000000    0.0000000000000000
     0.0000000000000000   13.0000000000000000    0.0000000000000000
     0.0000000000000000    0.0000000000000000   13.0000000000000000
   Zr   O   La   In   Li
   16   96   24    2   50
Direct
  0.7518668884963048    0.2491843318299140    0.7531853410886375
  0.2477259845414701    0.7494191872042241    0.2501502988711277
  0.7524691061644625    0.7494371570468047    0.2564397831666353
  0.2559149858739283    0.2486735688657034    0.7537520658110528
  0.2480365912082891    0.7443199362920460    0.7483090868935210
  0.7512989601678153    0.2414003528411754    0.2453937471484826
  0.2529110096503540    0.2489023870576836    0.2504251029537619
  0.7464349076061012    0.7505901257318174    0.7520560286674781
  0.4922884754959145    0.4917624930560456    0.0031993463763471
 -0.0039435824805122   -0.0057039157250749    0.5011019961410914
  0.5035773062705123   -0.0092813239071987    0.4973417598879536
 -0.0023354284939581    0.4965356781051097   -0.0016181791727322
  0.0056981546062684    0.4928565096402130    0.5025555626674647
  0.5035681260665738   -0.0074316135295295    0.0026192679370554
 -0.0139850537335926    0.0028000635145504    0.0065859145265004
  0.4932365573839582    0.4950671548165402    0.5063670831207303
  0.5574500518107239    0.3483887265827348    0.0421556453996564
  0.0605594268842246    0.8484562188865432    0.5425801124504336
  0.9496564614865201    0.6406033108797787    0.5423525925849217
  0.4432253925002407    0.1393671518760749    0.0279563459800191
  0.4446290191660247    0.8404121673826254    0.4669794227207621
  0.9426642975989696    0.3486793117769503    0.9630853905957388
  0.0441565941437753    0.1522670231473828    0.9697968201649226
  0.5470962428161910    0.6461932722474569    0.4715727497193652
  0.3517873146275841    0.0319991827831580    0.5497090109118002
  0.8501601731799918    0.5304251965107666    0.0607060602798812
  0.6423185807417129    0.5331647190664669    0.9460440751763924
  0.1488641950516392    0.0328165072323779    0.4429906027518660
  0.8502421026735619    0.4610144926231295    0.4473535530160488
  0.3477993843659634    0.9576354338592277    0.9414788385095951
  0.1507606780725366    0.9662261785932610    0.0437092338994469
  0.6527108638095822    0.4654381581229646    0.5476781216092362
  0.0321770117576373    0.5534235691344241    0.3557937718716715
  0.5352253671146990    0.0424161959916584    0.8488659850440456
  0.5218809328255352    0.9320964980385922    0.6488927677576831
  0.0335648999553937    0.4423811403113716    0.1521782291958538
  0.4741390039346972    0.4319420951313208    0.8536193151093251
  0.9752782197588101    0.9347250506142358    0.3521058071858080
  0.9732136943145180    0.0519250715261228    0.1567310194202681
  0.4614342525441235    0.5479990728192021    0.6546872412470193
  0.9487114929033446    0.1481486961540699    0.4710190793322047
  0.4416185797965771    0.6450263352296937    0.9689114167487583
  0.5500116690720400    0.8356116964036202    0.9727857864279920
  0.0599488748732834    0.3412932438928674    0.4753604151816228
```

```
 0.0483155572498523   0.6491913999533796   0.0276014999956189
 0.5557235306089516   0.1418064092008793   0.5300608301399778
 0.4424776084505125   0.3407802469646182   0.5274554355188540
 0.9419472998998931   0.8463010137182277   0.0289434325373295
 0.1482146055878910   0.4576464784292548   0.9526631284870948
 0.6488422134334699   0.9570485814562242   0.4398457940425884
 0.8483823707655914   0.9748720611615896   0.5612730847449526
 0.3415508437785135   0.4730640323360868   0.0642542165715520
 0.6504652121072435   0.0168626747131508   0.0552083364210547
 0.1550650065357669   0.5276653679996361   0.5582594097324386
 0.3469071740564853   0.5319580059618298   0.4503995680206980
 0.8437919015206592   0.0341945742052481   0.9430906928788444
 0.4660284883720570   0.9295704658559095   0.1473916676259609
 0.9658721021703102   0.4372098990318792   0.6515891140470786
 0.9706138255597144   0.5453706188919976   0.8441308946258265
 0.4683284610467119   0.0439717706949819   0.3469409294220315
 0.0293623162502986   0.0477506358924708   0.6542441478223214
 0.5262554453222565   0.5467954848077328   0.1552599842843242
 0.5291377470920510   0.4417197050983023   0.3516522830052096
 0.0277655606775781   0.9457607212860977   0.8509468049333824
 0.3024599143036314   0.7710439945659987   0.0976988065025237
 0.8018796182827270   0.2741430522219529   0.5987916456920599
 0.6960591830109654   0.2779039450973229   0.4033798177723927
 0.1872038525862192   0.7771977930591665   0.9009853945159095
 0.1954854906162619   0.2062922654750479   0.6078218412259359
 0.6974027099684940   0.7059051789033439   0.1079594098968224
 0.8048820371227141   0.7062154686904530   0.9005505540338198
 0.3034451072247367   0.2087544788090930   0.4025692601921472
 0.1053351325208565   0.3031156874318776   0.7803250807789073
 0.5988292998090350   0.7951734659668267   0.2817799775090310
 0.3977155519093213   0.6925971100574542   0.2823690519627542
 0.9025275801176276   0.1922953592268310   0.7777497352799698
 0.6027351486415176   0.1875559950323026   0.2250095726115219
 0.0964802630062733   0.6929564529688511   0.7170060365925512
 0.8951887893016895   0.8095739774656029   0.7281674123548844
 0.4020882131966915   0.3067060472219413   0.2280507260203663
 0.7873548620549921   0.1005024081604016   0.3101750677723321
 0.2823833368909097   0.5974864146351198   0.8042073962850217
 0.2868138115168460   0.3925720991713608   0.6907922575645865
 0.7855782950349379   0.8911683904578660   0.1951128330684355
 0.2084506352004590   0.6000352864528340   0.1962048654510977
 0.7099475625015086   0.0991540639807620   0.7070951213910546
 0.7095382214704451   0.8947245045661102   0.8152938527647250
 0.2132612376808108   0.3959077517811595   0.3082217536090326
 0.6984055124283821   0.2152632008127691   0.9043061512104505
 0.1886033753720894   0.7250118363714234   0.4016499621930950
 0.2955790652126609   0.7201935580352388   0.5985279013687637
 0.8125071273239212   0.2287948626409645   0.1000385486235748
 0.8082723492437106   0.7698976614112241   0.4066098311002466
 0.3144636320007951   0.2689543496484189   0.9007399350542785
 0.1945167910488325   0.2810496768648387   0.0995929618359532
 0.6882722907798913   0.7747347652560608   0.6017948752357097
 0.9065265569018653   0.6927380175066238   0.2221852460651932
 0.4087476094038182   0.1904107244333365   0.7191615742204736
 0.6017951381356235   0.3037296693773309   0.7138751769243864
 0.0985054361185395   0.8052301025286087   0.2123564663883464
 0.3992155783694867   0.7960830982124772   0.7841805210865103
 0.9025386381940929   0.3014760889874683   0.2870265262157057
 0.0966302830088486   0.1966211324917376   0.2822872350162506
 0.5931266215214814   0.6932169681998047   0.7828016476108168
 0.2200501484053940   0.9057003167389122   0.7099030681901685
 0.7183003872514236   0.4035438559482661   0.2106734524854819
 0.7159856081676371   0.5935106177985755   0.3141324414080294
 0.2145946589824770   0.0963716958549964   0.8127824963760707
 0.7757582701082630   0.4014259280304597   0.8124375355934884
 0.2855932544751775   0.9007122451768584   0.3063849811857202
 0.2754122206777675   0.0977274063853573   0.1971222509438522
 0.7792651500512295   0.5973434017902042   0.6998100466936940
 0.5033387961700843   0.2451788733753993   0.8886348341884308
-0.0021987747452957   0.7493682266755957   0.3846656733411913
 0.5001220270218681   0.7464090534776706   0.6259250257545167
-0.0004368823821114   0.2466776615460964   0.1271326735290509
 0.2493011692742123   0.8742513709462620   0.4974652839436348
 0.7450745408417927   0.3728655969879239   0.0010339282274596
 0.7571296026928477   0.6341560559202131   0.5132886865683854
 0.2592518053545174   0.1359744062415737   0.0114302405649778
 0.8879500879302006   0.4994967117366977   0.2473802147189044
 0.3884173429268070   0.0001827774343220   0.7462191707232648
```

```
  0.6215093377373989   0.4957291521177865   0.7563746216892815
  0.1241027223598643  -0.0022602867728115   0.2532353524802744
  0.0001827862098375   0.2436741677182074   0.6291076563681335
  0.4995581834100470   0.7421530476034952   0.1277837552440040
 -0.0062305031108218   0.7443116967624493   0.8709988923457990
  0.4982181418901296   0.2396561046776397   0.3709197243481133
  0.2492237446206008   0.6208827758085450   0.0018318924589972
  0.7496197299747135   0.1125485111445967   0.5002630037930771
  0.7492322933708652   0.8673117438043583   0.0046720646654116
  0.2511752996414590   0.3627685933481899   0.5022460706106114
  0.6295758861324231  -0.0031131714703157   0.2486373260807442
  0.1318240976135292   0.4911715736316138   0.7540539902246663
  0.8694123305820803  -0.0029145450507483   0.7479159855698951
  0.3693789723701927   0.4974113703361871   0.2503936920675427
  0.1894194874865732   0.9405913672028162   0.8799142413629747
  0.6913308657367968   0.4393522996496012   0.3795191272900885
  0.2067219040052106   0.5838562984653179   0.4252659915058750
  0.7588209254230324   0.8732249249571377   0.5058534335035648
  0.2720511662603406   0.3740443401776086  -0.0164086574814572
  0.0617857014416683   0.1309570024219620   0.8232871492588372
  0.3867681390970419   0.0128773409409633   0.2270775040460705
 -0.0996893626523601   0.6826009296881572   0.0742670370188529
  0.5573841464339057   0.8400683514770813   0.8297074324262381
  0.0571894366882852   0.3350751938500995   0.3302216629187312
  0.7456897822585810   0.6180044036638844   0.0029477261464287
  0.2447614762053526   0.1202605468969829   0.5084601434257707
  0.4339851771385884   0.6525731100432537   0.8237095124591851
  0.8723109944555957  -0.0074893506463132   0.2572198403748433
  0.1247865266165726   0.5019400685214030   0.2614580210780229
  0.4336514842855773   0.8524805207426847   0.3196499493401500
 -0.0198530193834949   0.7556544616235675   0.6261836707075614
  0.3217561522398197   0.0607558171948184   0.3727279345234314
  0.8230298363540725   0.5576221866502410   0.8420163886391033
  0.8151871970087365   0.4208454293275805   0.6020446360856796
  0.3021310773749981   0.9126098232626682   0.0813015032306761
  0.6711023290521682   0.5589768281314552   0.1687289683099272
  0.1716201236141702   0.0605256012867553   0.6700708878587444
  0.9461992239408178   0.8378277100367952   0.1758633077890475
  0.4534570023521242   0.3318868048134969   0.6735126349379742
  0.9368105808280845   0.1567062721379108   0.3273556363899625
  0.5674110681942043   0.1606171117354469   0.6754230980519538
  0.0626091238210339   0.6618911813491339   0.1723899124917680
  0.9019257880507100   0.1969846594788443   0.9227493350366388
  0.3988415773635690   0.6856354722856542   0.4284885458524020
  0.5932610461634656   0.8043738694208196   0.4257914734309694
  0.0825265547507108   0.3046524708741081   0.9221142107991598
  0.1365701190984286   0.6777829658227159   0.5720730883611876
  0.7471484823138886   0.1235919561894826   0.0249031740745834
  0.3099029107819762   0.5731811672934312   0.5975413128581112
  0.1765524710794325   0.4253622649425306   0.1119139835574389
  0.6300017777516534   0.9798064244474080   0.7265812234663528
  0.6817999925171871   0.0666569962388464   0.8702150395862842
  0.3723642993698661   0.4847210764423437   0.7647252706765865
  0.5636524942575530   0.6354365774885334   0.3248166377344197
  0.4959745540126313   0.2336938378335661   0.1279167115419434
  0.9248457918657276   0.5774122464032043   0.6896817444032097
  0.0732057417289859   0.9040670111792978   0.6893103902241410
  0.5688308187078782   0.3987667333714960   0.1902862359789771
  0.9241974832699353   0.3929207760630997   0.8102477730540197
  0.6467420027603165   0.3150842011254698   0.5647819700665846
  0.1184981875575225   0.8166302808331621   0.0626416849338774
  0.0937171959820056   0.1801566360604850   0.4269352941505293
  0.5875437627970475   0.6799139167164782   0.9272307629912947
  0.4091180531499832   0.1794348689599146   0.5717969533149364
  0.3906062926243097   0.8074542077954220   0.9331309525904214
  0.8717690236369892   0.3086507860566827   0.4371804030108298
CONTCAR_In_tet_051
llzo
 1.00000000000000000
     13.1340000000007997     0.0000000000000000     0.0000000000000000
      0.0000000000000000    13.1340000000007997     0.0000000000000000
      0.0000000000000000     0.0000000000000000    12.6630000000000003
     Zr   O   La   In   Li
     16   96   24    2   50
Direct
  0.2468921259161837   0.2571429181036339   0.2513374017202264
  0.7469373479254920   0.7570869881035996   0.7535826848367108
```

```
0.7566389654369141   0.2538590629662038   0.7523081248727981
0.2482268756836175   0.7564603028508351   0.2504536814444838
0.9833569059594908   0.4983624202081299   0.0032857374116675
0.4860830925216545   0.0035263624099473   0.5038001579161623
0.9963445968476962   0.5068858538421330   0.4994531412907647
0.4963656735197560   0.0063284955985918   0.9982129296481367
0.2476279535683841   0.2540823264930464   0.7503022174423339
0.7433068968639416   0.7583275216909242   0.2527915722075397
0.7487199567169960   0.2652591391162953   0.2487301986363155
0.2489555087910406   0.7624009088892398   0.7484685744062570
0.9966083642075997   0.0070555907456899   0.4998793877453946
0.4989903992598732   0.5055081869028719   0.0001377454665354
0.9951178643159230   0.0086997632642503   0.9981698558033616
0.4971533591647002   0.5088266525640925   0.4971645078371401
0.3029232630498554   0.2939262914499169   0.4048225481601245
0.8019009359570021   0.7904931604564713   0.9076012742825837
0.6877918931433812   0.2913139807760597   0.8988669247165884
0.1927984808523294   0.7882670866667687   0.4042991233052451
0.9678525770915569   0.4486617772861566   0.1569549840542725
0.4584058145151506   0.9521879540252619   0.6562487101018224
0.0306876134921414   0.4578493396718531   0.6529770955141319
0.5286692035727156   0.9511682972792435   0.1524304105322595
0.6917259308483649   0.7243288514517580   0.5979088864486051
0.1934689833320324   0.2231246629078908   0.0971583488389909
0.3038072433076293   0.7215242244473500   0.0964156167947590
0.8073806802517044   0.2292999788662656   0.5954231014245518
0.0242277999962912   0.5661891783470016   0.8487237023909819
0.5234131698587696   0.0617318206621671   0.3456963949918408
0.9593631918033393   0.5637354397649560   0.3451608502751976
0.4615927529995596   0.0645753051821337   0.8458486282225421
0.3022077639528265   0.2239184743606819   0.9047817285187898
0.8005153379460991   0.7240551201470303   0.4059058796822315
0.6920539742573862   0.2201441720728209   0.4094487317169000
0.1932847848220074   0.7235852634968042   0.9059598449774888
0.9751846335073092   0.0698772186145114   0.6481533067109299
0.4620330640533828   0.5610311116661586   0.1526606794095705
0.0345643652284979   0.0587991604189539   0.1551921568329595
0.5322447227205137   0.5635506634082006   0.6528272272232799
0.6888265034877821   0.7927950724497205   0.0978370592242207
0.1922170731507706   0.2939246579673359   0.5976805790215701
0.3003173027018083   0.7915076786438616   0.5966159146302655
0.8058188584022490   0.2823420621084392   0.1018530889235054
0.0348065901363140   0.9492452537309168   0.3444504268532255
0.5300822696386687   0.4521639509199497   0.8438901122677133
0.9600191462796348   0.9479457347943718   0.8498815901028113
0.4601410620037937   0.4525157690866110   0.3460521075227428
0.0988565295739845   0.2149877604601921   0.7906665319611484
0.5940940051711709   0.7113188701438072   0.2872655344755913
0.8935920694855444   0.2055897746399520   0.2874481068046556
0.3952997272489434   0.7024212024884562   0.7828930370671645
0.0495964837876762   0.6584439144773192   0.5349020526307739
0.5469157696739921   0.1543526144798912   0.0363766427964330
0.9338753497581697   0.6489264855075977   0.0400072682111756
0.4348031030157951   0.1527574589858889   0.5374657958041947
0.8907422799951223   0.8162575571758018   0.2156859129542192
0.3966947703344211   0.3123961020494920   0.7168387811231925
0.0988526217446930   0.8089178558727994   0.7167315755304468
0.5965940177990318   0.3158138451135695   0.2305103652073754
0.9454601736278135   0.3593116379805194   0.4644143277621834
0.4375654339677264   0.8588860399524594   0.9632962605591943
0.0503955893094292   0.3563543160911531   0.9667583563082242
0.5371726070494433   0.8554141635992311   0.4679847942261204
0.1021696543365226   0.3148728974478867   0.2869104767058176
0.5987322581984384   0.8127593368215699   0.7856290578873111
0.9022376123914563   0.3176746284246638   0.7737033178366072
0.3952950291767830   0.8149073606544398   0.2846828374204303
0.0505924386671284   0.8572029379226336   0.0348120600790775
0.5492341074336266   0.3607413676527497   0.5356974374845275
0.9394529756375858   0.8579659166376259   0.5229620612113455
0.4372059909819527   0.3601570696307642   0.0339919034629309
0.8932032694412653   0.6999841791422807   0.7146407194995346
0.3972558043097447   0.2064991113361477   0.2142673927725153
0.1004595754173646   0.7029700204878030   0.2129409616593909
0.6069587687926710   0.1965923103923586   0.7103562055474323
0.9417100933874648   0.1556421514496558   0.9812253472178134
0.4349816319361194   0.6538297470632598   0.4662602501441532
0.0528609163486370   0.1518521178393200   0.4584446531348690
0.5441745111106429   0.6557776827128160   0.9637378640954258
```

```
0.2683914670733134   0.1055324574811937   0.6964545042014850
0.7670297627731484   0.6106120112672824   0.2017495153088842
0.7141185840009159   0.1058136914589117   0.2055962474362303
0.2154561675507131   0.6040836655235630   0.7045881178414930
0.1474529099816847   0.4837970391008975   0.4409374480408651
0.6474208157938420   0.9793575474507211   0.9471013769553628
0.8401536715056753   0.4782209716889310   0.9400873849098061
0.3391745689641296   0.9785690924904752   0.4457346269426978
0.7120501235751370   0.9104446279452345   0.3106012303067136
0.2150929079374482   0.4106638766514580   0.8049331104639857
0.2721629604868773   0.9092957567591416   0.8032383052955990
0.7724241368859024   0.4080930031936414   0.3160441177457940
0.8439231128466590   0.5337312203493239   0.5515505171773873
0.3434277249555380   0.0352243477202955   0.0517725604829348
0.1457479189773505   0.5356023493921614   0.0437529471336388
0.6441352336350067   0.0298222031834050   0.5408853484272463
0.2738466993784224   0.4098466522062731   0.1951866900674386
0.7727271531835770   0.9086301608629384   0.7004939978308401
0.7186094257695805   0.4069354439685203   0.6996137333425448
0.2220814663063590   0.9063299702351852   0.1922618424800348
0.1461552105930740   0.0371648711935735   0.9457135580961046
0.6461307959556547   0.5386328225079973   0.4467502678392787
0.8451816434984021   0.0349140629837695   0.4335933149507346
0.3413726557587666   0.5358492714073204   0.9412599900582350
0.7172118131588059   0.6065613322008752   0.8049028184984446
0.2226426856350483   0.1076164381181557   0.3053053411271100
0.2751874250681821   0.6052580664891437   0.3028943672504097
0.7765088360468475   0.1054154308857582   0.8137094249501255
0.8436130163605882   0.9789823842954040   0.0539317195795841
0.3458130841122443   0.4807626136033801   0.5571387965448031
0.1462367960233775   0.9748750041356867   0.5512831083374363
0.6450473703894627   0.4809911914770757   0.0522023367252926
0.4922068365204009   0.2602901151400427   0.3772460577315472
0.9896497222486300   0.7628761659525870   0.8745687601394482
0.5033859835827654   0.2574145767671214   0.8764933610968885
0.9960316613808037   0.7552251933942515   0.3732680385480051
0.9929352724740016   0.2587911119832229   0.1266792940942822
0.4976551101416612   0.7576348866536095   0.6260914185697742
0.9950875820620563   0.2541391772557383   0.6245329092082457
0.4964493694360271   0.7569250101025744   0.1243705892971510
0.2429781163958344   0.1312216450782752   0.4998459417941781
0.7472527082998384   0.6261026280251951   0.0027470034120979
0.7489444211981264   0.1397255780978661   0.9997321562082871
0.2460909679246094   0.6272878493883843   0.4984024838002735
0.1223664230245649   0.5073592990178473   0.2494672444927930
0.6293270129835119   0.0103323206591926   0.7514862008651132
0.8699134579092402   0.5047430220040652   0.7490495568377183
0.3658402198775844   0.0063886709853642   0.2477053328348909
0.7590746495731115   0.8808749214420247   0.5134943335931917
0.2497043359508916   0.3746399092762904   0.0064869457592519
0.2456733234093637   0.8845877738139055   0.0009719312643577
0.7461536058135056   0.3829244200012686   0.5051440263291366
0.1127376024095385   0.0066018182445945   0.7505586715301694
0.6279591257699049   0.5021495166142865   0.2475393006838695
0.8801764847154314   0.0040526168550995   0.2433947660459860
0.3790865672633228   0.5028291837263826   0.7439949436962361
0.6842869498213230   0.0627688251811215   0.3720550167189162
0.1830188342575111   0.5631457163575549   0.8741986211125138
0.4996360356891141   0.2525477122008739   0.6247287050019452
0.9920818889525892   0.7537175319106483   0.1261339011640683
0.5089806760756971   0.2667921489051250   0.1234560327505882
0.9887890039201674   0.7700619267700144   0.6279791972001199
0.0005201446057605   0.2579143974273581   0.3732101145134077
0.4968380684727731   0.7605916323416048   0.8727848175203827
0.9767681982255672   0.2483111300531402   0.8784294644107428
0.4888217200762268   0.7514735386224709   0.3775899221326188
0.3110733413831550   0.0774761572748111   0.8856342807992659
0.8115295101735387   0.5757074232654282   0.3848133406726687
0.1785615030221195   0.0785953574271689   0.1115218251356265
0.6778794990985357   0.5781495589623317   0.6227942201644574
0.3157927407447159   0.5753227441203598   0.1243689867785511
0.3068482932325563   0.9408653841410189   0.6276465489771432
0.7501534927334087   0.3801315679043941   0.0119350564769309
0.6761384895589441   0.9413058551524955   0.1202704744493263
0.1744453820704812   0.4399618032096420   0.6427510473854150
0.8116771412664144   0.9398211162616744   0.8958892248639636
0.3136312898850553   0.4400930978688128   0.3701049807715251
0.1791223069040867   0.9347125678363066   0.3889577429036886
```

```
  0.6769473508298351   0.4380497587982749   0.8521471376614773
  0.3469763477813620   0.1787680691582237   0.0611620169867053
  0.8374125058737810   0.6750614752726037   0.5597272052825600
  0.6742648272522379   0.1691937939051032   0.5660857344854290
  0.1548027887799067   0.6803802307726211   0.0624562125346570
  0.0672117883960762   0.3807084527321464   0.8116203721862424
  0.5639285072783335   0.8949211968571046   0.3089890705149698
  0.9142995985989842   0.4193289855977069   0.3063317584336248
  0.4152619068660745   0.9236630687606309   0.8058798333859925
  0.6579139831045130   0.8358122287584762   0.9450091436367899
  0.1599282417256457   0.3415695944863743   0.4461651758325511
  0.3366031757355232   0.8348008923770949   0.4453982182013157
  0.9051393993963337   0.5805600808844911   0.1959685396994793
  0.4128589538783343   0.0940907775264397   0.6974238789815572
  0.0680099412818911   0.6076075168437485   0.6848963265402986
  0.5682079288224198   0.0989938834900724   0.1862916102372729
  0.3335558222779035   0.3381701376748445   0.5569943672711547
  0.8322440205247041   0.8354396770782491   0.0612329742058976
  0.1545648765757692   0.8310166631313785   0.5567742144307168
  0.0819684747783751   0.0896656187995122   0.3049103249080162
  0.5755458670148389   0.5986041719446906   0.8076429437601370
  0.4169786508118317   0.5910543178953089   0.3021242694432096
  0.6574371108237979   0.6817925279917598   0.4441665665228762
  0.1498071240267362   0.1825825769591548   0.9404020058660033
  0.3495899941664059   0.6825408921688840   0.9439687342575783
  0.8458759130530935   0.1810525904953575   0.4465950071480410
  0.8733579288743197   0.0165825360793228   0.7370729296024032
  0.4169634732838992   0.4180282281725241   0.1935170392216778
  0.0814839720358633   0.9172897436080533   0.1923710071112616
  0.5777271692646530   0.4155710299145478   0.6871903836251585

CONTCAR_LLZO_cub_049
llzo
 0.9985500000000000054
    13.000000000000000       0.0000000000000000       0.0000000000000000
     0.0000000000000000      13.000000000000000       0.0000000000000000
     0.0000000000000000       0.0000000000000000      13.000000000000000
   Zr   O   La   Li
    16   96   24   56
Direct
  0.7477496579271445   0.2489424055532025   0.7519996868655727
  0.2538408990534149   0.7520702089351778   0.2506154251833485
  0.7516888933599907   0.7567519567109586   0.2454241027457091
  0.2472955011858560   0.2538397988392599   0.7484446802723350
  0.2506782801996209   0.7491184414489491   0.7537956739463240
  0.7463062136664896   0.2529976733083431   0.2539624510566748
  0.2518959204072561   0.2538259136516811   0.2552786344054974
  0.7575305950623308   0.7520657625946636   0.7505223318514072
  0.5019385869721465   0.5013278057789796   0.0021676276867772
  0.0058709596632541   0.0034259834206488   0.5043332015793133
  0.5026669170957773   0.0004073033757465   0.5053340946735914
  0.9970153344671261   0.5027710303106717   0.0055584625693957
  0.0014718262447745   0.5049492493904520   0.4978704396602184
  0.4990684535449196   0.0074134860882002   0.9963653923436659
  0.0045038150310675   0.0056139169420846   0.0019174121526416
  0.5073274748610782   0.5034537596221507   0.4997499841896589
  0.5488602720161477   0.3487635025521972   0.0333721149927918
  0.0586524225249734   0.8494182713778996   0.5329760811030287
  0.9484308356321469   0.6583743693279299   0.5320000856328405
  0.4451483094745154   0.1585240165726428   0.0311739567448812
  0.4522236397545188   0.8447885186581587   0.4742485090736301
  0.9446389659443909   0.3504863486685746   0.9714970529552214
  0.0545076192145969   0.1605938456773708   0.9732894295950010
  0.5594058820130404   0.6562558479961824   0.4684152377865363
  0.3496022337532294   0.0249561532219139   0.5521939171525970
  0.8465672448247912   0.5292869426814865   0.0616985425650930
  0.6528753000344485   0.5286954424104883   0.9513359371119019
  0.1563248904553881   0.0378912016940429   0.4446326547400527
  0.8522464784399635   0.4809547968332240   0.4411955137005628
  0.3451338657447445   0.9736453081039876   0.9460279703307093
  0.1581065505480840   0.9757363880022939   0.0547494714604118
  0.6567832647738957   0.4727889581334232   0.5594157556602812
  0.0317828080323676   0.5592980489043582   0.3444723171976428
  0.5236120694264698   0.0596628565502494   0.8456941740611198
  0.5340544154560432   0.9479672349229005   0.6588385427072339
  0.0318190490968763   0.4490415678928286   0.1572027056989599
  0.4710212465241604   0.4520931267158791   0.8477991528631955
  0.9748721311204881   0.9505976452932663   0.3475713172595002
```

```
0.9724634694436308   0.0566422058616268   0.1553577849755716
0.4737871896110394   0.5576611126570971   0.6549488578033494
0.9511190231726754   0.1532674834566188   0.4635978617717348
0.4437059441945865   0.6518171604709490   0.9653795099484556
0.5601826296486486   0.8600848011168412   0.961143615380277
0.0566155552987702   0.3561373465636091   0.4600675421918946
0.0585464178907137   0.6505153918215233   0.0407863661356804
0.5611331859201709   0.1473712215095220   0.5399078823724383
0.4536191572634920   0.3555482582269159   0.5408385847964556
0.9453269377379356   0.8575126490393123   0.0371181825204464
0.1471117971282493   0.4657310046671145   0.9487655751509498
0.6508317932502191   0.9647327711323630   0.4441819400263077
0.8565047469725723   0.9633705947031952   0.5547461315972352
0.3518242529933653   0.4668442580374473   0.0606692883170232
0.6466124925733622   0.0403078963119939   0.0594068494683057
0.1508296464248334   0.5403943981040495   0.5561456206106625
0.3562911105196234   0.5413655371535382   0.4468064009270575
0.8563002902504981   0.0411867291163465   0.9462691793297507
0.4633320967235720   0.9447635268783061   0.1453569971243272
0.9728784484579541   0.4497870396060543   0.6487038974134250
0.9739837847060775   0.5611372031344729   0.8547077997095919
0.4662475424652279   0.0566964546228791   0.3542567582146521
0.0256534674255062   0.0643046299291236   0.6526296629231759
0.5372129828917424   0.5607717553425535   0.1515557729151480
0.5270105339098969   0.4421804629763267   0.3517718665835503
0.0401686128545684   0.9487978958425598   0.8532040360552942
0.3048610443938770   0.7822934232323133   0.0964456283902802
0.7974944285911725   0.2917894782781572   0.6010909930939606
0.6901456628174635   0.2750868843945379   0.4054667934158404
0.1981866614323139   0.7812207664057975   0.9074061682951818
0.1828836117034775   0.2363692630300868   0.6008069927154208
0.6977075530106679   0.7253532280387364   0.0912634527225126
0.8102911627446108   0.7179747047623383   0.9024124861852650
0.3000640454571382   0.2227879734167587   0.4076993188298028
0.0943119257936182   0.3087993197980338   0.7832461375180710
0.5968034299885455   0.8083609968219592   0.2773769321002176
0.4073190249398114   0.7008272679877589   0.2802697678646916
0.9020230294894208   0.2039149130667576   0.7835413773502570
0.5932914443411889   0.1980568799117396   0.2221820499441174
0.0984861680182783   0.6973544327642121   0.7236522696338348
0.9096092336697831   0.8055551467379761   0.7240754237325008
0.4045578251233592   0.3053395603349541   0.2232777573378694
0.7827500367696785   0.0994668261466838   0.3041374456353773
0.2803800756764325   0.5952941798836808   0.8067614838987525
0.2844161055728504   0.4067496558759695   0.6966751613638353
0.7799921603776314   0.9107390416956511   0.1946677378503770
0.2222549779604045   0.5990022824952028   0.1990396812519623
0.7287430213637546   0.0993201499079974   0.6880683578111043
0.7266530637421766   0.9022388242026812   0.8015013318712866
0.2219669795464554   0.4077999484792004   0.3061901770400710
0.6884319346181236   0.2282603609911247   0.9014581415108125
0.1948116392520572   0.7170950001365410   0.3983039684593907
0.3094300345842465   0.7161603678973234   0.6050467212024517
0.8015914136427528   0.2142894010929231   0.1022850201828207
0.8092358402008456   0.7893970815357703   0.3942672199779932
0.3049123047412154   0.2928615139315823   0.8985440235849640
0.1965782677188632   0.2906161606280013   0.1047232921171537
0.6973558187620972   0.7868566643217583   0.6019006978362649
0.8995131441947001   0.6987481227995329   0.2133084631546069
0.3925412848768158   0.1907545736516324   0.7238504447767289
0.5950463304112448   0.3044059851117253   0.7176076136185424
0.1053225022948634   0.8108675718316560   0.2136953692572865
0.3998490347425223   0.8044628585009920   0.7889951439026457
0.8950295212383181   0.3139472245791348   0.2755947227339495
0.1015486043853556   0.1938857097795507   0.2784077310204764
0.6080333184039122   0.6924958622965723   0.7891314858167040
0.2139987199870111   0.8975370989187340   0.6993265718121394
0.7089405645539563   0.3990688105302228   0.1918341053433601
0.7131933982521727   0.6092388254556591   0.3004477981930800
0.2107517764216403   0.1073885370913309   0.8119141354876275
0.7845329073726824   0.4030755279722839   0.8042515991474090
0.2904063473194683   0.9003817733824024   0.3101654392585524
0.2909371945841768   0.1054974472012356   0.1981437077398236
0.7884382059242238   0.60007051166754977   0.6969118800212148
0.4995826617119383   0.2525278288019562   0.8701011577055537
0.0009423054935256   0.7545181220460172   0.3687837974514770
0.5031497168409123   0.7520907231428814   0.6349548620138119
0.9997831307856809   0.2535543253771647   0.1344679394199237
```

```
    0.2507063692889212  0.8748123826339214  0.5046019482498666
    0.7495798819917047  0.3736280479987254  0.0025836603844548
    0.7516891601341471  0.6275001225563835  0.4979740766705867
    0.2504151584573028  0.1374450599777795  0.0007273306245388
    0.8669127325882475  0.5020895762132074  0.2521803899102407
    0.3673445322613786  0.0060229249666842  0.7494612163666077
    0.6293893580649418  0.5033117258682995  0.7529681121820893
    0.1335165222843246  0.0042037190081181  0.2520487955500172
    0.9981099280696006  0.2561287701724454  0.6250739540876564
    0.5013206784831554  0.7550716590953230  0.1168458317588811
    0.0016720677082257  0.7518950896961435  0.8821749426025965
    0.5009290236227796  0.2520137942324462  0.3861491671397857
    0.2495389734160669  0.6199556172435057  0.0029130722418490
    0.7520783290645049  0.1262504694988421  0.5016878290685792
    0.7498019951647857  0.8884975696107523  0.9962635016617053
    0.2524845764058293  0.3792718801416372  0.5032207740706218
    0.6199561243171816  0.0039540132067220  0.2501357693192021
    0.1197652573424332  0.5014777280901015  0.7530378489002620
    0.8781316929642885  0.0059846906735303  0.7542684257361804
    0.3828754617729997  0.5028628547003273  0.2528009337450457
    0.7433054090970328  0.1254891659219648  0.9867417057984766
    0.2524686193647693  0.6290151496942957  0.5014431729556517
    0.6827927104177481  0.9363769935177695  0.6534956331029176
    0.2499361410992968  0.3788906462046597  0.0012295809079925
    0.0637146992055976  0.8443828461019458  0.6829880768887429
    0.8279665219703918  0.4438266801556254  0.6601003307722396
    0.3786024320384622  0.0002005732077972  0.2484906897084918
    0.0001385620381516  0.7538307319257871  0.1252740584882326
    0.5078876166965848  0.7544946908338502  0.8759173055366436
    0.9385168561278162  0.2893052108855010  0.4058495095134731
    0.1877533360959469  0.9315877620036082  0.9026807543245569
    0.6846547708740525  0.5745451483473298  0.1035933703803607
    0.2437445183000906  0.1255342450479887  0.5371795836369924
    0.4329574572422861  0.4079815741378437  0.6865225074928704
    0.8265927602403854  0.9449922807846071  0.3432973397511180
    0.5652836453153832  0.6618001266240136  0.3183776322645254
    0.0625643927173787  0.1607191331446972  0.8258191547171346
    0.5603718625835153  0.3452816108687241  0.1801838977410689
    0.3210160974539455  0.0760817028867378  0.3973231263228484
    0.8303033972959457  0.5585445618891480  0.8472094712906166
    0.7571970399329851  0.3825848964595239  0.4809739598016851
    0.3136482466554421  0.9304735050017691  0.0899246841552839
    0.9273105434755979  0.9137399215716177  0.1862557644467723
    0.4806420433118658  0.2358578145047908  0.6230409065088610
    0.0733262631857949  0.4054782976003143  0.3138459057429941
    0.5839151792456446  0.9255868692908875  0.8138311477731012
    0.4269031799176183  0.5982900131542388  0.8178923308023087
    0.9334222727576846  0.1030823587595999  0.3154249380441608
    0.0737753490934888  0.5947505515706225  0.1911681092888747
    0.9089658918094325  0.1961273170241680  0.9303958887329373
    0.4120902037629344  0.6897800512007572  0.4288881196527430
    0.4050749340078260  0.3168151729642997  0.0756760319440806
    0.9139787708853810  0.8102079261713909  0.5779602028654117
    0.5973281265326418  0.8143888301923909  0.4272861911170153
    0.0895669938674135  0.3179204437727730  0.9284078614284249
    0.0916065070146485  0.6897047777728922  0.5770825883576922
    0.5966079817773197  0.1917752678885007  0.0741620261994793
    0.3258352824816452  0.5631332267225173  0.6575259422560904
    0.8235574719750973  0.0705235071467605  0.1540753279622042
    0.1788348132751104  0.4390778675453884  0.1588775375654981
    0.5839135180584878  0.0891315183467316  0.7030670927248803
    0.1784741888684313  0.5648948915275622  0.3454050069986852
    0.7554749341449067  0.8757089787666464  0.4985401468420880
    0.3253011695204885  0.4392367124271014  0.8413351766647102
    0.1274693807953734  0.0099916370250341  0.7397809069380675
    0.4386851970155014  0.1623233671614994  0.1785437232397276
    0.9351322175592615  0.6275463250137464  0.6827040329641636
    0.6288236696582487  0.4950916593762726  0.2656668130431656
    0.9301421156315719  0.3630909926250645  0.8225618421117751
    0.4398716920277308  0.8452818738121813  0.3266755823825969
    0.6312607665441113  0.3245448053162599  0.5772848043490027
    0.1595831487016669  0.8271612694757159  0.0632200367727041
    0.1049279641624100  0.1914371836203691  0.4239004981044031
    0.6716718687163735  0.6715132911200447  0.9381961719876767
    0.8377510318513540  0.6744212781090767  0.0510964733347000
    0.3446794730850016  0.8248761503973838  0.9364457984350797
```

CONTCAR_LLZO_tet

```
llzo
   1.00000000000000000
     13.1340000000000003    0.0000000000000000    0.0000000000000000
      0.0000000000000000   13.1340000000000003    0.0000000000000000
      0.0000000000000000    0.0000000000000000   12.6630000000000003
    Zr   O   La   Li
    16   96   24   56
Direct
  0.2453270000000032  0.2570089999999965  0.2501960000000025
  0.7453270000000032  0.7570089999999965  0.7501960000000025
  0.7453270000000032  0.2570089999999965  0.7501960000000025
  0.2453270000000032  0.7570089999999965  0.2501960000000025
  0.9953270000000032  0.5070089999999965  0.0001960000000025
  0.4953270000000032  0.0070089999999965  0.5001960000000025
  0.9953270000000032  0.5070089999999965  0.5001960000000025
  0.4953270000000032  0.0070089999999965  0.0001960000000025
  0.2453270000000032  0.2570089999999965  0.7501960000000025
  0.7453270000000032  0.7570089999999965  0.2501960000000025
  0.7453270000000032  0.2570089999999965  0.2501960000000025
  0.2453270000000032  0.7570089999999965  0.7501960000000025
  0.9953270000000032  0.0070089999999965  0.5001960000000025
  0.4953270000000032  0.5070089999999965  0.0001960000000025
  0.9953270000000032  0.0070089999999965  0.0001960000000025
  0.4953270000000032  0.5070089999999965  0.5001960000000025
  0.3010958600417382  0.2915124584845594  0.4040419849096253
  0.8010958600417381  0.7915124584845596  0.9040419849096251
  0.6895591399582658  0.2915124584845594  0.9040419849096251
  0.1895591399582657  0.7915124584845596  0.4040419849096253
  0.9608235415154400  0.4512401399582615  0.1540419849096253
  0.4608235415154402  0.9512401399582616  0.6540419849096251
  0.0298304584845591  0.4512401399582615  0.6540419849096251
  0.5298304584845592  0.9512401399582616  0.1540419849096253
  0.6895591399582658  0.7225055415154404  0.5963490150903754
  0.1895591399582657  0.2225055415154404  0.0963490150903752
  0.3010958600417382  0.7225055415154404  0.0963490150903752
  0.8010958600417381  0.2225055415154406  0.5963490150903754
  0.0298304584845591  0.5627768600417339  0.8463490150903754
  0.5298304584845592  0.0627768600417340  0.3463490150903751
  0.9608235415154400  0.5627768600417339  0.3463490150903751
  0.4608235415154402  0.0627768600417340  0.8463490150903754
  0.3010958600417382  0.2225055415154406  0.9040419849096251
  0.8010958600417381  0.7225055415154404  0.4040419849096253
  0.6895591399582658  0.2225055415154406  0.4040419849096253
  0.1895591399582657  0.7225055415154404  0.9040419849096251
  0.9608235415154400  0.0627768600417340  0.6540419849096251
  0.4608235415154402  0.5627768600417339  0.1540419849096253
  0.0298304584845591  0.0627768600417340  0.1540419849096253
  0.5298304584845592  0.5627768600417339  0.6540419849096251
  0.6895591399582658  0.7915124584845596  0.0963490150903752
  0.1895591399582657  0.2915124584845594  0.5963490150903754
  0.3010958600417382  0.7915124584845596  0.5963490150903754
  0.8010958600417381  0.2915124584845594  0.0963490150903752
  0.0298304584845591  0.9512401399582616  0.3463490150903751
  0.5298304584845592  0.4512401399582615  0.8463490150903754
  0.9608235415154400  0.9512401399582616  0.8463490150903754
  0.4608235415154402  0.4512401399582615  0.3463490150903751
  0.0974193098674893  0.2016383415378014  0.7850209340727115
  0.5974193098674891  0.7016383415378014  0.2850209340727113
  0.8932356901325076  0.2016383415378014  0.2850209340727113
  0.3932356901325075  0.7016383415378014  0.7850209340727115
  0.0506976584622054  0.6549166901325105  0.5350209340727115
  0.5506976584621983  0.1549166901325104  0.0350209340727113
  0.9399563415378009  0.6549166901325105  0.0350209340727113
  0.4399563415378009  0.1549166901325104  0.5350209340727115
  0.8932356901325076  0.8123796584621987  0.2153710659272866
  0.3932356901325075  0.3123796584621988  0.7153710659272865
  0.0974193098674893  0.8123796584621987  0.7153710659272865
  0.5974193098674891  0.3123796584621988  0.2153710659272866
  0.9399563415378009  0.3591003098674922  0.4653710659272866
  0.4399563415378009  0.8591003098674921  0.9653710659272865
  0.0506976584622054  0.3591003098674922  0.9653710659272865
  0.5506976584621983  0.8591003098674921  0.4653710659272866
  0.0974193098674893  0.3123796584621988  0.2850209340727113
  0.5974193098674891  0.8123796584621987  0.7850209340727115
  0.8932356901325076  0.3123796584621988  0.7850209340727115
  0.3932356901325075  0.8123796584621987  0.2850209340727113
  0.0506976584622054  0.8591003098674921  0.0350209340727113
  0.5506976584621983  0.3591003098674922  0.5350209340727115
```

```
0.9399563415378009  0.8591003098674921  0.5350209340727115
0.4399563415378009  0.3591003098674922  0.0350209340727113
0.8932356901325076  0.7016383415378014  0.7153710659272865
0.3932356901325075  0.2016383415378014  0.2153710659272866
0.0974193098674893  0.7016383415378014  0.2153710659272866
0.5974193098674891  0.2016383415378014  0.7153710659272865
0.9399563415378009  0.1549166901325104  0.9653710659272865
0.4399563415378009  0.6549166901325105  0.4653710659272866
0.0506976584622054  0.1549166901325104  0.4653710659272866
0.5506976584621983  0.6549166901325105  0.9653710659272865
0.2727975447341431  0.1058225261258792  0.6963787829249601
0.7727975447341430  0.6058225261258794  0.1963787829249602
0.7178574552658538  0.1058225261258792  0.1963787829249602
0.2178574552658537  0.6058225261258794  0.6963787829249601
0.1465134738741204  0.4795384552658566  0.4463787829249602
0.6465134738741203  0.9795384552658567  0.9463787829249601
0.8441415261258836  0.4795384552658566  0.9463787829249601
0.3441415261258833  0.9795384552658567  0.4463787829249602
0.7178574552658538  0.9081944738741161  0.3040122170750403
0.2178574552658537  0.4081944738741164  0.8040122170750403
0.2727975447341431  0.9081944738741161  0.8040122170750403
0.7727975447341430  0.4081944738741164  0.3040122170750403
0.8441415261258836  0.5344785447341459  0.5540122170750403
0.3441415261258833  0.0344785447341460  0.0540122170750403
0.1465134738741204  0.5344785447341459  0.0540122170750403
0.6465134738741203  0.0344785447341460  0.5540122170750403
0.2727975447341431  0.4081944738741164  0.1963787829249602
0.7727975447341430  0.9081944738741161  0.6963787829249601
0.7178574552658538  0.4081944738741164  0.6963787829249601
0.2178574552658537  0.9081944738741161  0.1963787829249602
0.1465134738741204  0.0344785447341460  0.9463787829249601
0.6465134738741203  0.5344785447341459  0.4463787829249602
0.8441415261258836  0.0344785447341460  0.4463787829249602
0.3441415261258833  0.5344785447341459  0.9463787829249601
0.7178574552658538  0.6058225261258794  0.8040122170750403
0.2178574552658537  0.1058225261258792  0.3040122170750403
0.2727975447341431  0.6058225261258794  0.3040122170750403
0.7727975447341430  0.1058225261258792  0.8040122170750403
0.8441415261258836  0.9795384552658567  0.0540122170750403
0.3441415261258833  0.4795384552658566  0.5540122170750403
0.1465134738741204  0.9795384552658567  0.5540122170750403
0.6465134738741203  0.4795384552658566  0.0540122170750403
0.4953270000000032  0.2570089999999965  0.3751960000000025
0.9953270000000032  0.7570089999999965  0.8751960000000025
0.4953270000000032  0.2570089999999965  0.8751960000000025
0.9953270000000032  0.7570089999999965  0.3751960000000025
0.9953270000000032  0.2570089999999965  0.1251960000000025
0.4953270000000032  0.7570089999999965  0.6251960000000025
0.9953270000000032  0.2570089999999965  0.6251960000000025
0.4953270000000032  0.7570089999999965  0.1251960000000025
0.2453270000000032  0.1292408580432722  0.5001960000000025
0.7453270000000032  0.6292408580432721  0.0001960000000025
0.7453270000000032  0.1292408580432722  0.0001960000000025
0.2453270000000032  0.6292408580432721  0.5001960000000025
0.1230951419567275  0.5070089999999965  0.2501960000000025
0.6230951419567275  0.0070089999999965  0.7501960000000025
0.8675598580432692  0.5070089999999965  0.7501960000000025
0.3675598580432693  0.0070089999999965  0.2501960000000025
0.7453270000000032  0.8847761419567305  0.5001960000000025
0.2453270000000032  0.3847761419567303  0.0001960000000025
0.2453270000000032  0.8847761419567305  0.0001960000000025
0.7453270000000032  0.3847761419567303  0.5001960000000025
0.1230951419567275  0.0070089999999965  0.7501960000000025
0.6230951419567275  0.5070089999999965  0.2501960000000025
0.8675598580432692  0.0070089999999965  0.2501960000000025
0.3675598580432693  0.5070089999999965  0.7501960000000025
0.4953270000000032  0.2570089999999965  0.6251960000000025
0.9953270000000032  0.7570089999999965  0.1251960000000025
0.4953270000000032  0.2570089999999965  0.1251960000000025
0.9953270000000032  0.7570089999999965  0.6251960000000025
0.9953270000000032  0.2570089999999965  0.3751960000000025
0.4953270000000032  0.7570089999999965  0.8751960000000025
0.9953270000000032  0.2570089999999965  0.8751960000000025
0.4953270000000032  0.7570089999999965  0.3751960000000025
0.6762692109556930  0.0760667890443066  0.3751960000000025
0.1762692109556931  0.5760667890443066  0.8751960000000025
0.3143857890443107  0.0760667890443066  0.8751960000000025
0.8143857890443108  0.5760667890443066  0.3751960000000025
```

```
0.1762692109556931  0.0760667890443066  0.1251960000000025
0.6762692109556930  0.5760667890443066  0.6251960000000025
0.8143857890443108  0.0760667890443066  0.6251960000000025
0.3143857890443107  0.5760667890443066  0.1251960000000025
0.3143857890443107  0.9379502109556889  0.6251960000000025
0.8143857890443108  0.4379502109556889  0.1251960000000025
0.6762692109556930  0.9379502109556889  0.1251960000000025
0.1762692109556931  0.4379502109556889  0.6251960000000025
0.8143857890443108  0.9379502109556889  0.8751960000000025
0.3143857890443107  0.4379502109556889  0.3751960000000025
0.1762692109556931  0.9379502109556889  0.3751960000000025
0.6762692109556930  0.4379502109556889  0.8751960000000025
0.3329039351145971  0.1772284755049033  0.0565528159946329
0.8329039351146045  0.6772284755049034  0.5565528159946328
0.6577500648853948  0.1772284755049033  0.5565528159946328
0.1577500648853949  0.6772284755049034  0.0565528159946329
0.0751075244950964  0.4194320648853955  0.8065528159946328
0.5751075244950963  0.9194320648853952  0.3065528159946329
0.9155464755049030  0.4194320648853955  0.3065528159946329
0.4155464755049028  0.9194320648853952  0.8065528159946328
0.6577500648853948  0.8367895244950967  0.9438381840053677
0.1577500648853949  0.3367895244950969  0.4438381840053676
0.3329039351145971  0.8367895244950967  0.4438381840053676
0.8329039351146045  0.3367895244950969  0.9438381840053677
0.9155464755049030  0.5945859351146049  0.1938381840053676
0.4155464755049028  0.0945859351146047  0.6938381840053677
0.0751075244950964  0.5945859351146049  0.6938381840053677
0.5751075244950963  0.0945859351146047  0.1938381840053676
0.3329039351145971  0.3367895244950969  0.5565528159946328
0.8329039351146045  0.8367895244950967  0.0565528159946329
0.6577500648853948  0.3367895244950969  0.0565528159946329
0.1577500648853949  0.8367895244950967  0.5565528159946328
0.0751075244950964  0.0945859351146047  0.3065528159946329
0.5751075244950963  0.5945859351146049  0.8065528159946328
0.9155464755049030  0.0945859351146047  0.8065528159946328
0.4155464755049028  0.5945859351146049  0.3065528159946329
0.6577500648853948  0.6772284755049034  0.4438381840053676
0.1577500648853949  0.1772284755049033  0.9438381840053677
0.3329039351145971  0.6772284755049034  0.9438381840053677
0.8329039351146045  0.1772284755049033  0.4438381840053676
0.9155464755049030  0.9194320648853952  0.6938381840053677
0.4155464755049028  0.4194320648853955  0.1938381840053676
0.0751075244950964  0.9194320648853952  0.1938381840053676
0.5751075244950963  0.4194320648853955  0.6938381840053677
```